\documentclass[11pt]{amsart}
\usepackage[utf8]{inputenc}

\usepackage[T1]{fontenc}

\usepackage[ 
text={16cm,22cm},
headheight=9pt,
centering
]{geometry}

\usepackage{xcolor}
\definecolor{darkblue}{RGB}{0,0,160}
\usepackage{hyperref}
\hypersetup{
colorlinks,%
citecolor=black,%
filecolor=black,%
linkcolor=darkblue,%
urlcolor=darkblue
}

\usepackage{chngcntr}
\counterwithin{figure}{section}
\usepackage{latexsym,array,delarray,amsthm,amssymb,epsfig,amsmath,blkarray,tikz}%eufrak
\usepackage{url}
\usetikzlibrary{decorations.markings}
\usetikzlibrary{arrows, automata}
\usepackage{enumitem}
\usepackage{xspace}

\theoremstyle{plain}
\newtheorem{thm}{Theorem}[section]

\newtheorem{prop}[thm]{Proposition}
\newtheorem{cor}[thm]{Corollary}

\newtheorem*{thm*}{Theorem}
\newtheorem*{lemma*}{Lemma}
\newtheorem*{prop*}{Proposition}
\newtheorem*{cor*}{Corollary}
\newtheorem*{conj*}{Conjecture}

\theoremstyle{definition}
\newtheorem{defn}[thm]{Definition}
\newtheorem{ex}[thm]{Example}

\theoremstyle{remark}
\newtheorem*{rmk}{Remark}

\newcommand{\calr}{\mathcal{R}}

\DeclareMathOperator{\rank}{rank}

\newcommand{\ind}{\mbox{$\perp \kern-5.5pt \perp$}}

\newcommand{\pa}{\mathrm{pa}}

\newcommand{\seth}[1]{\begingroup #1\endgroup}

\allowdisplaybreaks[1]

\title{Phylogenetic Network Models as Graphical Models}
\author{Seth Sullivant}
\email{smsulli2@ncsu.edu}
\address{Department of Mathematics, North Carolina State University, Raleigh, NC 27695}

\date{\today}

\begin{document}

\begin{abstract}
    The displayed tree phylogenetic network model is shown to sit
    as a natural submodel of the graphical model associated to a directed
    acyclic graph (DAG).
    This representation allows us  to derive a number of results about 
    the displayed tree model.  In particular, the concept
    of a local modification to a DAG model is developed and applied 
    to the displayed tree model.  As an application, some nonidentifiability
    issues related to the displayed tree models are highlighted as they
    relate to reticulation edges and stacked reticulations in the networks.
    We also derive rank conditions on flattenings of probability tensors
    for the displayed tree model, generalizing
    classic results for phylogenetic tree models.
\end{abstract}

\maketitle

\section{Introduction}

Phylogenetic trees are the basic object used to represent the evolutionary
relationships between a collection of taxa \cite{Felsenstein2003, Steel2016}.  
However, there are many situations
when a more complex network structure is necessary to describe the
history of a collection of species.   The network structure can take
into account hybridization and more general types of reticulate evolution.  

Once a particular network is specified, there are numerous models for 
how evolution might occur on that network.  The network multispecies coalescent
is a widely used model that allows for both the population level effect 
of incomplete lineage sorting, and for hybridization \cite{yu2012}.  It is usually
used as a process for generating gene trees, with a separate
substitution process of the evolution of sequences along those gene trees.
Questions about identifiability in the network multispecies coalescent have
often involved inferring the species network from gene trees, and there
are many different approaches to such identifiability results 
\cite{Allman2024,Allman2023}.  

An alternate family of network models consider evolution of sequences
on the tree contained within that network,
but without the coalescent process.  This model is sometimes
called the \emph{displayed tree model}, because the only gene trees that can be 
created are the trees that are displayed by the network.  
 Identifiability questions around the
displayed trees model have been addressed in a few papers \cite{GrossLong2018,Grossvan2021}.  Those works specifically address
the level-1 networks for group-based models.  
Note that the displayed tree model can be thought of as a limiting family of the
network multispecies coalescent, where all coalescent events occur
at the same time as speciation \cite{Rhodes2025}.

In this note we explain how to think of the displayed tree
phylogenetic network model as a natural submodel of a corresponding
graphical model on a directed acyclic graph (DAG).
DAG graphical models are a commonly used family of statistical models
that give a recursive factorization of a joint probability
distribution based on the underlying graph structure.
This representation  of the displayed tree model
as a DAG model is useful for proving some results about network
models by using properties of DAG models, like conditional
independence and the local structure of the conditional 
distributions in the model.  One feature we highlight here
is that the DAG structure often allows us to prove uniform
results across all model types (not just group-based models,
or the general Markov model).

One goal for this note is to make the connection between the 
displayed tree model and the DAG models more widely known.  
We use this fact to 
prove some straightforward results about the displayed tree model,
and we want to advertise this perspective in the hope that might be useful 
for proving other results about phylogenetic network models.  Along the way,
we see some surprising nonidentifiability results for
the displayed tree model, especially as it involves reticulation
vertices.

\begin{ex}
    As an example of results produced in this note, 
    consider the three networks given in Figure \ref{fig:stacked2}.
    For evolutionary models satisfying common mathematical assumptions
    (in particular, for the equivariant phylogenetic models), all three
    networks produce the same family of distributions on sequences at the
    leaves ($\alpha, \beta, \gamma, \delta$).  
    The sequences of reticulations are not
    the same in the left and right networks.  The issue that causes this phenomenon
    is the presence of stacked reticulations in the network.  Specifically, vertices $3$ and $5$
    are reticulations, and there is an edge from $3$ to $5$.
\end{ex}

\begin{figure}

 \begin{tikzpicture}[scale=1, every node/.style={circle, draw, fill=white, inner sep=2pt}]
  \tikzset{edge/.style = {->,> = latex}}
    \node (1) at (0,3) {1};
    \node (2) at (2,3) {2};
    \node (3) at (0,2) {3};
    \node (4) at (2,2) {4};
    \node (5) at (0,1) {5};
    \node (6) at (2,1) {6};
    \node (a) at (0,4) {$\alpha$};
    \node (b) at (2,4) {$\beta$};
    \node (c) at (2,0) {$\gamma$};
    \node (d) at (0,0) {$\delta$};
    
    \draw[edge] (1) to (a);
     \draw[edge] (1) to (3);
    \draw[edge] (2) to (1);
    \draw[edge] (2) to (b);
     \draw[edge] (2) to (4);
     \draw[edge] (4) to (3);
     \draw[edge] (3) to (5);
     \draw[edge] (4) to (6);
     \draw[edge] (6) to (5);
     \draw[edge] (5) to (d);
     \draw[edge] (6) to (c);
  \end{tikzpicture} \quad \quad \quad \quad
   \begin{tikzpicture}[scale=1, every node/.style={circle, draw, fill=white, inner sep=2pt}]
  \tikzset{edge/.style = {->,> = latex}}
    \node (1) at (0,3) {1};
    \node (2) at (2,3) {2};
    \node (4) at (2,2) {4};
    \node (5) at (0,1) {5};
    \node (6) at (2,1) {6};
    \node (a) at (0,4) {$\alpha$};
    \node (b) at (2,4) {$\beta$};
    \node (c) at (2,0) {$\gamma$};
    \node (d) at (0,0) {$\delta$};
    
    \draw[edge] (1) to (a);
    \draw[edge] (2) to (1);
    \draw[edge] (2) to (b);
     \draw[edge] (2) to (4);
     \draw[edge] (4) to (5);
     \draw[edge] (1) to (5);
     \draw[edge] (4) to (6);
     \draw[edge] (6) to (5);
     \draw[edge] (5) to (d);
     \draw[edge] (6) to (c);
  \end{tikzpicture} \quad \quad \quad \quad
     \begin{tikzpicture}[scale=1, every node/.style={circle, draw, fill=white, inner sep=2pt}]
  \tikzset{edge/.style = {->,> = latex}}
    \node (1) at (0,3) {1};
    \node (2) at (2,3) {2};
    \node (3) at (1,1)  {3};
    \node (4) at (2,2) {4};
    \node (5) at (0,1) {5};
    \node (6) at (2,1) {6};
    \node (a) at (0,4) {$\alpha$};
    \node (b) at (2,4) {$\beta$};
    \node (c) at (2,0) {$\gamma$};
    \node (d) at (0,0) {$\delta$};
    
    \draw[edge] (1) to (a);
    \draw[edge] (2) to (1);
    \draw[edge] (2) to (b);
     \draw[edge] (2) to (4);
     \draw[edge] (4) to (3);
     \draw[edge] (1) to (5);
     \draw[edge] (4) to (6);
     \draw[edge] (6) to (3);
      \draw[edge] (3) to (5);
     \draw[edge] (5) to (d);
     \draw[edge] (6) to (c);
  \end{tikzpicture}

\caption{Two binary networks (left and right) and one non-binary network (middle) that give the same distributions 
on the observed leaves
$\alpha, \beta, \gamma, \delta$. \label{fig:stacked2}}
  
\end{figure}

The outline of the paper is as follows.  Sections \ref{sec:graphical}
and \ref{sec:network} provide background on graphical models and 
the displayed tree phylogenetic network model and shows how
the displayed tree model arises as a submodel of directed acyclic
graph models (DAGs).
Section \ref{sec:localmod} introduces the notion of a local modification
in a DAG, as a tool to show that two DAGs give the same family of probability
distributions.  Section \ref{sec:stacked} explores the consequences of the local
modification machinery for the case of stacked reticulations.
This produces some situations where two network models produce the same
families of probability distributions in rather non-obvious ways.
In Section \ref{sec:nonidentcond}, some observations are made about
loss of identifiability in the conditional distributions associated
to reticulation vertices.  Finally, in Section \ref{sec:rankofflat}
conditional independence in DAGs is used to derive some results about the
ranks of flattenings of probability tensors in the case of the
displayed tree model.

%%%%%%%%%%%%%%%%%%%%%%%%%%%%%%%%%%%%%%%%%%%%%
%%%%%%%%%%%%%%%%%%%%%%%%%%%%%%%%%%%%%%%%%%%%%
%%%%%%%%%%%%%%%%%%%%%%%%%%%%%%%%%%%%%%%%%%%%%
\section{Background on graphical models} \label{sec:graphical}

This section reviews background on directed graphical models
that will be useful for studying the displayed tree model.
More details can be found in \cite{Lauritzen1996}.

Let $G = (V,E)$ be a directed acyclic graph with vertex set $V$.
The graph is directed acyclic, in that there are no directed
cycles in the graph, though there can be cycles that are not directed.
For each $i \in V$, let $\pa(i) = \{j \in V:  j \rightarrow i \in E \}$
be the set of parents of a node $i$.

For each $i \in V$, we have a random variable $X_i$.  Let $X = (X_i :  i \in V)$, be the random vector.  For a set $A \subseteq V$, let 
$X_A = (X_a : a \in A)$ be the subvector with indices indexed by $A$.
We assume that all of the random variables are discrete
so we can talk about probability distributions (but for continuous
random variables, we can use density functions instead).
So each random variable $X_i$ has a state space $[r_i] := \{1,2, \ldots, r_i\}$.
We let $\calr = \prod_{i \in V}  [r_i]$, be the state space
of the random vector.   For each $x \in \calr$, and $A \subseteq V$, 
we can also take subvectors $x_A$.  Let $\calr_A$ denote the state space
of the random vector $X_A$, that is, $\calr_A = \prod_{i \in A} [r_i]$.

Define $p(x) = P(X = x) $ to be the joint probability distribution of $X$.  In the event that $V = [n]$,
we can write this as
\[
p(x) = p(x_1, \ldots, x_n)  =   P(X_1 = x_1, \ldots, X_n = x_n).
\]
 For any $A \subseteq V$
we can compute the marginal distribution
\[
p_A(x_A)  =  P(X_A = x_A)  =  \sum_{y_B \in \calr_B}  p(x_A, y_B).
\]
For $A, B \subseteq V$, disjoint, we can compute the conditional distribution
\[
p_{A|B}(x_A|x_B)  =  P(X_A = x_A | X_B = x_B)  =  
\frac{p_{A \cup B}(x_A, x_B)}{p_B(x_B)}.
\]
In the special case that $A = \{i\}$ is a singleton, we just use $p_{i|B}(x_i | x_B)$.

Consider the following factorization of the joint probability according
to the DAG $G$
\begin{equation}\label{eq:factorization}
    p(x)  =  \prod_{i \in V}p_{i|\pa(i)}(x_i | x_{\pa(i)}).
\end{equation}
Note that not every probability distribution can satisfy this 
equation.  
%In general, one only has
%\[
%p(x)  =  \prod_{i \in [n]} p_{i| [i-1]}(x_i | x_{[i-1]}),
%\]
%which is sometimes called the multiplication rule for conditional
%probabilities.
The graphical model associated to the DAG $G$ consists of all 
probability distributions on $\calr$ that satisfy (\ref{eq:factorization}).

Alternatively, we can also think about (\ref{eq:factorization})
as a parametrization of a model, with each
of the conditional distributions $p_{i|\pa(i)}(x_i | x_{\pa(i)})$
as a set of free parameters.  Here we work with the constraint that they
should actually be conditional probability distributions,
that is
\begin{enumerate}
    \item $p_{i|\pa(i)}(x_i | x_{\pa(i)}) \geq 0$ and
    \item  $\sum_{x_i \in [r_i]} p_{i|\pa(i)}(x_i | x_{\pa(i)} ) = 1$.
\end{enumerate}

\begin{figure}
    \centering
 \begin{tikzpicture}[scale=1, every node/.style={circle, draw, fill=white, inner sep=2pt}]
  \tikzset{edge/.style = {->,> = latex}}
    \node (1) at (0,2) {1};
    \node (2) at (-1,1) {2};
    \node (3) at (1,1)  {3};
    \node (4) at (0,0) {4};

    \draw[edge] (1) to (2);
    \draw[edge] (1) to (3);
    \draw[edge] (2) to (4);
    \draw[edge] (3) to (4);
  
    \end{tikzpicture}
    \caption{A directed four-cycle $C_4$.}
    \label{fig:fourcycle}
\end{figure}
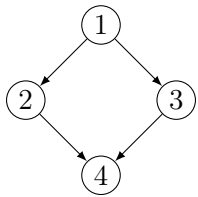

\begin{ex}
Consider the directed four-cycle graph $C_4$ in Figure \ref{fig:fourcycle}.
The parent sets of each of the vertices are $\pa(1) = \emptyset$,
$\pa(2) = \{1\}$, $\pa(3) = \{1\}$, $\pa(4) = \{2,3\}$.
The factorization of the joint distribution induced by $C_4$
 is:
\begin{equation}\label{eq:c4fac}
    p(x_1, x_2, x_3, x_4) =  p_1(x_1) p_{2|1}(x_2|x_1) p_{3|1}(x_3|x_1)
p_{4|2,3}(x_4 | x_2, x_3).
\end{equation}
The graphical model associated to $C_4$ consists of all probability distributions
that satisfy the factorization (\ref{eq:c4fac}).
\end{ex}

%%%%%%%%%%%%%%%%%%%%%%%%%%%%%%%%%%%
%%%%%%%%%%%%%%%%%%%%%%%%%%%%%%%%%%%
%%%%%%%%%%%%%%%%%%%%%%%%%%%%%%%%%%%
%%%%%%%%%%%%%%%%%%%%%%%%%%%%%%%%%%%

\section{The displayed tree model} \label{sec:network}

Phylogenetic network models arise as special cases of
the general DAG graphical model by putting multiple
types of restrictions on the DAGs that can arise,
the particular structure of the conditional distributions
that are used, and the fact that many of the variables are 
unobserved random variables (i.e.~hidden random variables or latent
random variables).
Throughout, all the random variables $X_i$ are assumed to have the
same state space $\Sigma$ 
(which for many models is $\Sigma = \{\mathtt{A},\mathtt{C},\mathtt{G},\mathtt{T}\}$).

First, we describe the restrictions on the DAGs that can arise.
Different types of vertices in a network model have different names:
\begin{itemize}
    \item Root vertices:  indegree 0
    \item Leaf vertices:  outdegree 0 
    \item Tree vertices:  indegree 1
    \item Reticulation vertices:  indegree  $> 1$
\end{itemize}
Typically (though not exclusively) we assume that there is a single
root in $G$.  This node represents a common ancestor of all
the nodes in the graph.  Sometimes the graphs that can arise in network
models are further restricted in various ways, but technically
any other type of directed acyclic graph could occur.

Often one restricts to binary phylogenetic networks, in which
case we assume that the root has degree $2$, leaves have degree $1$,
and all other vertices have degree $3$. 
However, we will see that even if we only care about binary
phylogenetic networks, it is useful to look at phylogenetic networks
more broadly.
There are many other families of network restrictions that one
could add which concern the relations between types of vertices
in the network.  See \cite{Kong2022} for an extensive
survey on different types of phylogenetic networks. 

In many modeling contexts with phylogenetic networks, it
is useful to consider networks with parallel edges.  However, for the displayed tree model
with a mutation process that is closed under convex combinations, this model is equivalent to a model
without parallel edges, so we can safely ignore this condition and assume our DAGs are simple.  
This will be discussed
in detail in  Proposition \ref{prop:2blob}.
For this note, we do not need to make any other
restrictions.  

\begin{figure}

  \begin{tikzpicture}[scale=1, every node/.style={circle, draw, fill=white, inner sep=2pt}]
  \tikzset{edge/.style = {->,> = latex}}
    \node (1) at (1,1.5) {1};
    \node (2) at (1,2.5) {2};
    \node (3) at (2,3)  {3};
    \node (4) at (3,2.5) {4};
    \node (5) at (3,1.5) {5};
    \node (6) at (2,1) {6};
    \node (a) at (0,1) {$\alpha$};
    \node (b) at (0,3) {$\beta$};
    \node (c) at (2,4) {$\gamma$};
    \node (d) at (4,3) {$\delta$};
    \node (e) at (4,1) {$\epsilon$};
    \node (f) at (2,0)  {$\zeta$};

    \draw[edge] (1) to (a);
    \draw[edge] (1) to (6);
    \draw[edge] (2) to (1);
    \draw[edge] (2) to (b);
    \draw[edge] (3) to (2);
    \draw[edge] (3) to (c);
    \draw[edge] (3) to (4);
    \draw[edge] (4) to (d);
    \draw[edge] (4) to (5);
    \draw[edge] (5) to (e);
    \draw[edge] (5) to (6);
    \draw[edge] (6) to (f);
    \end{tikzpicture}
\quad \quad 
\begin{tikzpicture}[scale=1, every node/.style={circle, draw, fill=white, inner sep=2pt}]
  \tikzset{edge/.style = {->,> = latex}}
    \node (1) at (1,1.5) {1};
    \node (2) at (1,2.5) {2};
    \node (3) at (2,3)  {3};
    \node (4) at (3,2.5) {4};
    \node (5) at (3,1.5) {5};
    \node (6) at (2,1) {6};
    \node (a) at (0,1) {$\alpha$};
    \node (b) at (0,3) {$\beta$};
    \node (c) at (2,4) {$\gamma$};
    \node (d) at (4,3) {$\delta$};
    \node (e) at (4,1) {$\epsilon$};
    \node (f) at (2,0)  {$\zeta$};

    \draw[edge] (1) to (a);
    \draw[edge] (1) to (6);
    \draw[edge] (2) to (1);
    \draw[edge] (2) to (b);
    \draw[edge] (3) to (2);
    \draw[edge] (3) to (c);
    \draw[edge] (3) to (4);
    \draw[edge] (4) to (d);
    \draw[edge] (4) to (5);
    \draw[edge] (5) to (e);
    % \draw[edge] (5) to (6);
    \draw[edge] (6) to (f);
    \end{tikzpicture}  \quad \quad 
\begin{tikzpicture}[scale=1, every node/.style={circle, draw, fill=white, inner sep=2pt}]
  \tikzset{edge/.style = {->,> = latex}}
    \node (1) at (1,1.5) {1};
    \node (2) at (1,2.5) {2};
    \node (3) at (2,3)  {3};
    \node (4) at (3,2.5) {4};
    \node (5) at (3,1.5) {5};
    \node (6) at (2,1) {6};
    \node (a) at (0,1) {$\alpha$};
    \node (b) at (0,3) {$\beta$};
    \node (c) at (2,4) {$\gamma$};
    \node (d) at (4,3) {$\delta$};
    \node (e) at (4,1) {$\epsilon$};
    \node (f) at (2,0)  {$\zeta$};

    \draw[edge] (1) to (a);
   % \draw[edge] (1) to (6);
    \draw[edge] (2) to (1);
    \draw[edge] (2) to (b);
    \draw[edge] (3) to (2);
    \draw[edge] (3) to (c);
    \draw[edge] (3) to (4);
    \draw[edge] (4) to (d);
    \draw[edge] (4) to (5);
    \draw[edge] (5) to (e);
    \draw[edge] (5) to (6);
    \draw[edge] (6) to (f);
    \end{tikzpicture}

\caption{A 6 sunlet network, and its two displayed trees. \label{fig:6sunlet}}

\end{figure}
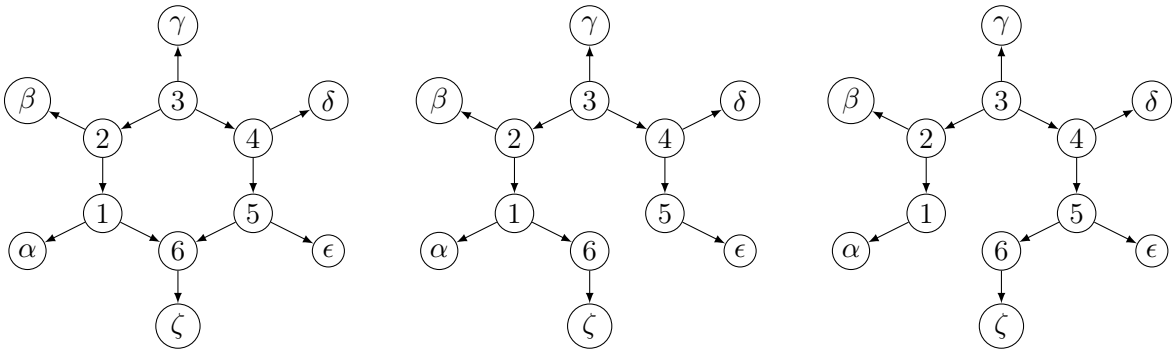

\begin{ex}
    Consider the network on the left in Figure \ref{fig:6sunlet}.
    The root is vertex $3$.  Vertices $\alpha, \beta, \gamma, \delta, \epsilon,$ and $\zeta$ are leaves.  Vertex $6$ is a reticulation vertex.
    All other vertices are tree vertices.  
\end{ex}

A second consideration when describing a network model, is identifying
a set of variables that are the observed variables.
In many network modeling scenarios, the leaves of the network are assumed to be the
only observed variables, though
 this assumption is not technically necessary.
However, we will assume that each observed node has no descendant
node that is hidden.  
This satisfies the natural assumption that observed variables should correspond
to extant taxa, i.e.~taxa that we can directly observe.
For a given set of observed taxa $O$ and hidden taxa $H$, the 
distribution of states at the observed variables is obtained by marginalizing
over $H$, that is
\[
p_O(x_O)  =  \sum_{x_H \in \calr_H}  p(x_O, x_H).
\]

Finally, we come to the model description which amounts to 
restrictions on the structure of the conditional
distributions $p_{i|\pa(i)}(x_i | x_{\pa(i)} ) $.
The restriction in the network displayed tree model
amounts to the following restrictions.

\begin{itemize}
    \item To each edge $e = i \to j$ in the network we associate a
    Markov transition matrix, which is the conditional distribution of
    $X_j | X_i$.  This is denoted $M^{ij}$, with entries $M^{ij}(x_j|x_i)$.  
    \item  The matrices $M^{ij}$ are usually not allowed to be arbitrary
    conditional distributions (as they would be in the graphical model),
    but rather are restricted to have a particular structure.
    Example structures include models like the Jukes-Cantor model,
    general Markov model, Kimura models, HKY model, general time reversible,
    etc.  Matrices might also be required to be of the form $\exp(Qt)$
    for some rate matrix $Q$ and branch length parameter $t$.
    \item  For each reticulation vertex $j$ (that is $\# \pa(j) > 1$),
    there is a probability vector $\pi^j \in \Delta^{\pa(j)}$. 
    The coordinate $\pi^j_i$ is the probability that the edge $i \to j$
    is chosen at the reticulation vertex.  Note if the non-root vertex $j$ 
    is not a reticulation vertex, 
    so it has only
    one parent $i$, then we can define $\pi^i_j = 1$.  This will be useful in some proofs.
    \item  With these features, we get the following formula for the
    conditional probability at each reticulation vertex $j$:
    \[
    p_{j|\pa(j)}(x_j | x_{\pa(j)}) =  \sum_{i \in \pa(j)} \pi^j_i  M^{ij}(x_j|x_i).
    \]
\end{itemize}

\seth{The displayed tree model is usually presented as follows:}  for each reticulation vertex $j$, 
one of the edges $ i \to j$ is chosen with probability $\pi^j_i$.
The resulting DAG obtained after making all these choices has no
reticulation vertices, and hence no cycles, so it is a
forest (and with appropriate assumptions on the
network, it will be  a rooted tree).  
The probabilities are then calculated according
to the graphical model on that forest as described above.
This clearly yields the same description of the model
as above using the graphical model formulation.
Indeed, this can be seen by using the graphical model
formulation, and using the distributive law to expand and
collect monomials in powers of the $\pi^j_i$ terms. 
Note that it suffices to see that the models are the same
when all variables are observed variables (since then this
will also be true when some of the variables are hidden).

\begin{ex}
Consider the four-cycle graph from Figure \ref{fig:fourcycle}.  
There is only
one reticulation vertex which is vertex $4$.  The parametrization
for the graphical model in this graph is
\[
p(x_1, x_2, x_3, x_4) =  p_1(x_1) p_{2|1}(x_2|x_1) p_{3|1}(x_3|x_1)
p_{4|2,3}(x_4 | x_2, x_3).
\]
In the displayed tree model, we parametrize $p_{4|2,3}(x_4 | x_2, x_3)$ as
\[
p_{4|2,3}(x_4 | x_2, x_3)  =  \pi^4_2 M^{24}(x_4|x_2) + \pi^4_3 M^{34}(x_4|x_3)
\]
and 
\[
p_{2|1}(x_2|x_1) = M^{12}(x_2|x_1) \quad  
p_{3|1}(x_3|x_1) = M^{13}(x_3|x_1)   
\]
so we get 
\begin{eqnarray*}
p(x_1, x_2, x_3, x_4) &= &  \quad  p_1(x_1) p_{2|1}(x_2|x_1) p_{3|1}(x_3|x_1)
p_{4|2,3}(x_4 | x_2, x_3) \\
& = &\quad  p_1(x_1) M^{12}(x_2|x_1) M^{13}(x_3|x_1) (\pi^4_2 M^{24}(x_4|x_2) + \pi^4_3 M^{34}(x_4|x_3) ) \\
& = & \quad \pi^4_2 p_1(x_1) M^{12}(x_2|x_1) M^{13}(x_3|x_1) M^{24}(x_4|x_2) \\
& &  + \pi^4_3 p_1(x_1) M^{12}(x_2|x_1) M^{13}(x_3|x_1) M^{34}(x_4|x_3). 
\end{eqnarray*}
The first term in this final sum is $\pi^4_2$ times the 
probability distribution in the tree obtained by deleting the edge $3 \to 4$, and the second term is $\pi^4_3$ times the probability
distribution in the tree obtained by deleting the edge $2 \to 4$.
\end{ex}

\begin{ex}
    For a more phylogenetics relevant example, consider the 
    network in Figure \ref{fig:6sunlet} on the left.  This has just one
    reticulation vertex which is vertex $6$.  There are two displayed trees
    (shown in the center and on the right), which are each obtained by deleting
    one of the reticulation edges ($1 \rightarrow 6$ or $5 \rightarrow 6$).
\end{ex}

\subsection{Equivariant phylogenetic models}

There are a number of classes of algebraic models that are useful to study,
because they occur in phylogenetics practice or because they have
nice mathematical properties, or both.
Among the most useful classes to study for mathematical reasons
in phylogenetics are the equivariant phylogenetic models.
These are phylogenetic models where there is a certain group
symmetry that is present among the states of the random
variables.  See \cite{Casanellas2023} for more
details on equivariant models. 
Specifically, let $\Sigma$ denote the set of states
of the random variables (e.g.~ for DNA sequences, $\Sigma = \{{\tt A}, {\tt C}, {\tt G}, {\tt T} \}$).
Let $\mathcal{G}$ be a group acting on $\Sigma$.  A set of transition matrices $TM$
is called equivariant relevant to $\mathcal{G}$
if for all $M \in TM $, all $x_1, x_2 \in \Sigma$ and all $g \in \mathcal{G}$
\[
M( x_2 | x_1)  =   M( g(x_2) | g(x_1)).
\]
The model is a \emph{generic equivariant model} relevant to the particular group
action if it consists of all transition matrices that satisfy the 
given symmetry conditions.  The \emph{open generic equivariant model} consists
of all transition matrices that satisfy the given equivariant condition
but with no zeroes in the transition matrices.  Adding the
open condition can be useful if we want to impose the condition
that a network is never allowed to have zero branch lengths (which
correspond to the transition matrix being the identity matrix).  

Examples of equivariant models include the Cavendar-Farris-Neyman model (CFN), 
the Jukes-Cantor model (JC), the Kimura 2 and 3 parameter models (K2P, K3P),
the strand symmetric model (SSM), and the general Markov model (GMM).   In general,
we use, $TM$, to denote a set of transition matrices that define our model.  Once $TM$
is specified, the displayed tree model associated to a particular DAG $G$ 
consists of all probability distributions that can arise on the observed variables,
as the transition matrices range over all possible values in $TM$.

\begin{ex}
    Let $\Sigma = [k]  = \{1, \ldots,  k \}$, and let $\mathcal{G} = \langle 1 \rangle$ be the 
    trivial group.  Then the corresponding equivariant model is the general
    Markov model, consisting of all $k \times k$ transition matrices.  Note that this
    set of transition matrices has $k(k-1)$ free parameters, since there are no restrictions on
    it besides being a transition matrix.
\end{ex}

\begin{ex}
    Let $\Sigma = \{{\tt A}, {\tt C}, {\tt G}, {\tt T} \}$, and let $\mathcal{G}$ be the symmetric group $S_4$
    acting by permuting this set of size four.  The corresponding general
    equivariant model is the Jukes-Cantor model consisting of all transition
    matrices of the form
    \[
\begin{pmatrix}
    1-3a & a & a & a \\
    a & 1-3a & a & a \\
    a & a & 1-3a & a \\
    a & a & a & 1-3a
\end{pmatrix}.
        \]
        \end{ex}

\subsection{Closure properties of sets of transition matrices}

In this subsection, we consider ways that the set $TM$ of
transition matrices could be closed under various operations.

\begin{defn}
    Let $TM$ be a set of transition matrices.  The set $TM$ is
    \emph{closed under multiplication} if for all $M^1, M^2 \in TM$,
    $M^1 M^2 \in TM$.    The set $TM$ is \emph{closed under convex combinations}
    if for all $M^1, M^2 \in TM$ and $\delta \in [0,1]$, 
    $\delta M^1 + (1- \delta) M^2 \in TM$.
\end{defn}

A more complicated property, that we will need for a result about $2$-blobs,
concerns convex combinations of marginal distributions.

\begin{defn}\label{defn:closedmarginal}
    Let $TM$ be a set of transition matrices and $RD$ be a set of corresponding root distributions
    in the transition model.  Consider the set of two variable marginal distributions
    \[
    MD =  \{  (M(x_b|x_a) \pi(x_a))_{x_a, x_b \in \Sigma}  :  M \in TM,  \pi \in RD \}.
    \]
    Then $TM$ is \emph{closed under marginal convex combinations}, if for all $\delta \in [0,1]$
    and all $P^1, P^2 \in MD$,  $\delta P^1 + (1-\delta)P^2 \in MD$.
\end{defn}

It is straightforward to see that general equivariant models 
(and general open equivariant models) satisfy
all three of these properties.

\begin{prop}
    Let $TM$ be a set of transition matrices from a general equivariant
    phylogenetic model.  Then $TM$ is closed under multiplication, convex
    combinations, and marginal convex combinations.  This is also true
    if $TM$ is the open general equivariant phylogenetic model.
\end{prop}

\begin{proof}
Being closed under multiplication of matrices follows immediately from the equivariant property.  This is because the equivariant property says that the
matrices are invariant under conjugation by certain permutation matrices,
which holds after multiplying the matrices together.  

Being closed under both types of convex combinations follows because the models are the intersection of a linear space with a convex set.  
\end{proof}

The closure properties can be more subtle in the case where we are working with
a set of transition matrices that is not as nice as a general equivariant model.
For instance, if we take the set of transition matrices
\[
TM  =  \{  \exp(Qt)  :  Q \in RM,  t \in [0, \infty)  \}
\]
where $RM$ is some collection of rate matrices.  Such a set is generally
not convex, and need not be closed under matrix multiplication.
One model of this type that does satisfy all three properties
is the \emph{continuous time} Jukes-Cantor model.  Let
\[
Q^{JC}  =  \begin{pmatrix}
    -3 & 1 & 1 & 1  \\
    1 & -3  & 1  & 1 \\
    1 & 1 & -3 & 1  \\
    1 & 1 & 1 & -3
\end{pmatrix}.
\]

\begin{prop}
Let $TM =  \{   \exp(Q^{JC}t)  :  t \in [0, \infty)  \} $ or 
$TM =  \{   \exp(Q^{JC}t)  :  t \in (0, \infty)  \}. $ 
Then $TM$ is closed under multiplication, convex
combinations, and marginal convex combinations.  
\end{prop}

\begin{proof}
    Clearly  
    \[
    \exp(Q^{JC}t_1)\exp(Q^{JC}t_2)  =  \exp(Q^{JC}(t_1 +t_2)) 
    \]
    so $TM$ is closed under matrix multiplications.

    Both $TM$ and $MD$ in this case are convex sets with
    \[
TM  =   \left\{  
\begin{pmatrix} 1-3a & a & a & a \\
a & 1-3a  & a & a \\
a & a & 1-3a & a  \\
a & a & a & 1-3a
\end{pmatrix}:    a \in [0, 1/4) \right \}
    \]
and
\[
MD = \left\{ \frac{1}{4} 
\begin{pmatrix} 1-3a & a & a & a \\
a & 1-3a  & a & a \\
a & a & 1-3a & a  \\
a & a & a & 1-3a
\end{pmatrix}   : a \in [0, 1/4) \right \}
    \]
(and for the case where we leave out $t = 0$, remove the identity matrix).  
So the models will be closed under convex combinations and marginal
convex combinations.
\end{proof}

\subsection{Splittability}

Besides being closed under products of transition matrices and
convex combinations, we will also be interested in sets of
transition matrices that have another
useful property we call splittability.  Splittability is
a feature of a set of transition matrices that will allow us to prove,
in some instances, that two networks produce the same probability distributions.
In a sense, splittability is like the inverse property to saying the family of
distributions is closed under products.  

\begin{defn}
    Let $TM$ be a set of transition matrices.  The set $TM$ is called \emph{splittable}
    if for any $k$ and any  $M^1, \ldots, M^k \in TM$, there is a 
    transition matrix $N \in TM$ such that $M^1N^{-1}, \ldots, M^k N^{-1}  \in TM$.
\end{defn}

Clearly if $TM$ contains the identity matrix, then it will 
be splittable, since we can take $N$ to be the identity matrix. 

\begin{prop}
    If $TM$ is a general equivariant phylogenetic model, then
    $TM$ is splittable.
\end{prop}

\begin{proof}
    The general equivariant phylogenetic model contains the identity
    matrix, so in the definition of splittable we can always take $N$ to
    be the identity matrix.
\end{proof}

However, we are sometimes
interested in restricting to sets of transition matrices that
exclude the identity (since this corresponds to branch length zero), or
excludes transition matrices which have zeros in them (as in the open
general equivariant models, described above).
Even in this case, there are often situations where the set of transition
matrices is still splittable.

\begin{prop}
    Let $TM$ be the set of transition matrices of a general open equivariant phylogenetic
    model (that is all entries of the transition matrices are positive).
    Then $TM$ is splittable.
\end{prop}

\begin{proof}
    Let $M^1, \ldots, M^k \in TM$.  Let $\| \cdot \|$ be a 
    submultiplicative matrix norm (that is, $\|A B \| \leq \|A \|  \| B \|$).
    Let $B_\delta(M) =  \{A :  \| M - A \| < \delta \}$ be the open ball of radius $\delta$ centered at $M$ with respect to the matrix norm and confined to the linear
    space of matrices with the given equivariant structure.  Choose $\epsilon > 0$ 
     smaller than 
    \[
     \min_{i = 1}^k  \sup \{  \delta : B_\delta(M^i) \subseteq TM \} .
    \]
    By the conditions on our matrices,  $\epsilon $ is strictly positive. 
    Let
    \[
    \mu = \max( 1, \max_{i = 1}^k  \| M^i \| ).  \]

Let $N$ be any matrix in $TM$ such that
\[
\| I - N \|  < \frac{\epsilon}{2 \mu} .
\]
Note that if $A = I-N$, then
\[
N^{-1}  =  \sum_{n = 0}^\infty  A^n
\]
so that
\[
\| N^{-1} \|  \leq   \frac{1}{1 - \|A \| }  \leq 2  .
\]
since we assume that $\|A \|$ is very small.
Then
\[
\| I - N^{-1} \| =  \|N^{-1} ( N - I) \|  \leq \|N^{-1} \|  \|N- I \| \leq  \frac{\epsilon}{\mu}.
\]

Then for each $i$ we have that 
\[
\| M^i  -  M^iN^{-1}\|  =  \| M^i (I - N^{-1}) \| \leq \|M^i\| \| I - N^{-1} \| 
<  \mu \cdot  \frac{\epsilon}{\mu}  =  \epsilon.
\]
So $M^i N^{-1}$ is a matrix of the correct equivariant form, and it has distance
$< \epsilon $ to $M^i$.  By our assumption on $\epsilon$, $B_\epsilon(M^i) \subseteq TM$
so all the entries of $M^i N^{-1}$  are positive,
and $M^i N^{-1} \in TM$ as desired.  This shows that $TM$ is splittable. 
\end{proof}

%%%%%%%%%%%%%%%%%%%%%%%%%%%%%%%%%%%%%
%%%%%%%%%%%%%%%%%%%%%%%%%%%%%%%%%%%%%
%%%%%%%%%%%%%%%%%%%%%%%%%%%%%%%%%%%%%
%%%%%%%%%%%%%%%%%%%%%%%%%%%%%%%%%%%%%

\section{Local modifications to DAGs}  \label{sec:localmod}

In this section, we want to discuss a useful property for looking at
local features in a graphical model, and using that to compare if two graphical models
describe the same family of probability distributions
by looking at local features of the graph.

\begin{defn}
      Let $G = (V,D)$ be a DAG.  Let $A, B, C$ be disjoint subsets of $V$ with the
    following properties:
\begin{itemize}
    \item  For each vertex $b \in B$, every edge $i \rightarrow b$ has $i \in A \cup B$
    \item  For each vertex $b \in B$,  every edge $b \rightarrow i$ has $i \in B \cup C$
    \item  For each vertex $c \in C$, every edge $i \rightarrow c$ has $i \in A \cup B \cup C$.
    \item  For each vertex $c \in C$, there is no directed path starting at $c$ and ending at a vertex
    $a \in A$.
\end{itemize}
We say that the triple of sets of vertices $(A,B,C)$ gives a \emph{local structure} in $G$.
\end{defn}

\begin{figure}
    \centering
    \includegraphics[width=0.5\linewidth]{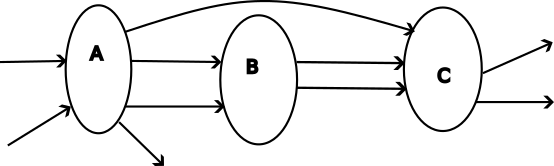}
    \caption{The diagram gives the idea of a local structure.  Note that there can be
    directed edges within each of the groups $A$, $B$, and $C$.  }
    \label{fig:localstructure}
\end{figure}

With these conditions on our graph, and the subsets $A,B,C$, we can directly
calculate the conditional probability of $X_C$ given $X_A$ in the DAG $G$ by the following
formula:

\[ 
p_{C|A}(x_C | x_A)  =   \sum_{x_B \in \calr_B}  \prod_{i \in B \cup C } 
p_{i|\pa(i)} ( x_i | x_{\pa(i)} ).
\]

To prove this formula, it suffices to prove that
\[ 
p_{B \cup C|A}(x_B, x_C | x_A)  =   \prod_{i \in B \cup C } 
p_{i|\pa(i)} ( x_i | x_{\pa(i)} )
\]
since the previous formula is obtained by marginalization of the second formula.

Let $S$ be the ancestral set of $A \cup B \cup C$, that is, $S$ consists of all
vertices  $s \in V$ such that there is a path from $s$ to some $t \in A \cup B \cup C$.  A basic fact about sets $S$ that are ancestral is that the joint distribution
for probabilities in $S$ has the form
\[
p_S(x_S)  =  \prod_{i \in S}  p_{i|\pa(i)}(x_i| x_{\pa(i)})
\]
which holds because for each $i \in S$ each of the parent sets $\pa(i) \subseteq S$ 
when $S$ is an ancestral set.
Note that if $(A,B,C)$ is a local structure for $G$, then the ancestral set $S'$ for
$A$, is simply $S'  =  S \setminus (B \cup C).$
Thus we have
\[
p_{S'}(x_{S'})  =  \prod_{i \in S'} p_{i | \pa(i)} (x_i | x_{\pa(i)}).
\]
From this we can see that 
\begin{eqnarray*}
p_{B \cup C | S'}( x_B, x_C| x_{S'}) & = &  \frac  {\prod_{i \in S}  p_{i|\pa(i)}(x_i| x_{\pa(i)})} {\prod_{i \in S'} p_{i | \pa(i)} (x_i | x_{\pa(i)} )  }\\
& = &  \prod_{i \in B \cup C } 
p_{i|\pa(i)} ( x_i | x_{\pa(i)} ).
\end{eqnarray*}
However, since for each $i \in B \cup C$, our assumptions of being a local
structure imply that each set $\pa(i) \subseteq A \cup B$.  In particular, this shows that
the final formula only depends on elements in $A \cup B \cup C$, so that 
\[
p_{B \cup C | S'}( x_B, x_C| x_{S'})  =  p_{B \cup C | A}( x_B, x_C| x_{A}).
\]

The fact that the conditional distribution for a local structure is local
only to the variables $A,B,C$ means that we can compare two graphical models
that only differ by a change in a local structure.  We call this a 
local modification.

\begin{defn}
    Let $G$ be a DAG with a local structure $(A,B,C)$.  Let $V' = V \setminus (A \cup B \cup C)$  Let $G'$ be a new DAG with vertex set $V' \cup A \cup B' \cup C$
    that satisfies the following properties 
    \begin{itemize}
        \item $(A,B',C) $ is a local structure in $G'$.
        \item Let $i,j \in V' \cup A$.  Then $i \rightarrow j \in G$ if and only if $i \rightarrow j \in G'$.
        \item  Let $i \in C$ and $j \in V'$.  Then $i \rightarrow j \in G$ if and only if $i \rightarrow j \in G'$.
     \end{itemize}
     The graphs $G$ and $G'$ are called \emph{local modifications} of each other.
\end{defn}

In summary, graphs $G$ and $G'$ are local modifications of each other if they 
are exactly the same graph outside of the local structures $(A,B,C)$ and $(A,B',C)$.
Note that in the "exactly the same" category we are requiring that any
edge $a \rightarrow a' \in G$ also appears in $G'$ .

\begin{ex}
    The most basic example of a local modification is to subdivide an edge.
    Specifically, if $a \rightarrow c$ is any edge in $G$.  We can take
    $A = \{a\},  B = \emptyset,  C = \{c\}$ in $G$.  In $G'$ we replace the
    edge $a \rightarrow c$ with the pair of edges $a \rightarrow b$ and
    $b \rightarrow c$ where $b$ is a new vertex and take $B' = \{b\}$.
    Then $G$ and $G'$ are local modifications of each other.  This
    particular type of local modification is the subject of
    Proposition \ref{prop:degree2vertex}.
 \end{ex}

From the standpoint of comparing displayed tree models on different graphs, 
there are some situations where two graphs that are local modifications
of each other can yield the same family of probability distributions.
The structure of a local modification means that we can check
this condition by purely looking at the induced structure of the conditional
distributions in the two changed substructures.

Note that we use notation like $p_{G, A}(x_A)$ and $p_{G, A|C}(x_A | x_C)$,
when we need to refer to the distributions that specifically come from the
graph $G$.

\begin{thm}\label{thm:local}
  Let $G = (V, D)$ and $G' = (V', D') $ be two DAGS that are local modifications of each other
    with local structures $(A,B,C)$ and $(A,B',C)$ respectively.
    Suppose that the family of conditional distributions in the two models
    $p_{G, C|A}(x_C|x_A)$ and $p_{G', C|A}(x_C|x_A)$ are the same.
    Suppose further that each of the other families of  distributions $p_{i|\pa(i)}(x_i, x_{\pa(i)})$
    is the same in both models.
    Then the family of joint distributions with the variables in $X_B$ and $X_{B'}$
    hidden variables are the same in both models.
\end{thm}

\begin{proof}
    The distributions with $B$ and $B'$ hidden in both models looks like
  \begin{eqnarray*}
      p_{G}(x_{V \setminus B})  & = &
      \sum_{x_B \in \calr_B} \prod_{i \in V} p_{i| \pa(i)}(x_i| x_{\pa(i)})  \\
      & = &  \prod_{i \in V \setminus (B \cup C)} p_{i| \pa(i)}(x_i| x_{\pa(i)} ) 
      \sum_{x_B \in \calr_B} \prod_{i \in B \cup C} p_{i| \pa(i)}(x_i| x_{\pa(i)} )
  \end{eqnarray*}  
  and
  \begin{eqnarray*}
      p_{G'}(x_{V' \setminus B'})  & = &
      \sum_{x_{B'} \in \calr_{B'}} \prod_{i \in V'} p_{i| \pa(i)}(x_i| x_{\pa(i)})  \\
      & = &  \prod_{i \in V' \setminus (B' \cup C)} p_{i| \pa(i)}(x_i| x_{\pa(i)} ) 
      \sum_{x_{B'} \in \calr_{B'}} \prod_{i \in B' \cup C} p_{i| \pa(i)}(x_i| x_{\pa(i)} ).
  \end{eqnarray*}  

These factorizations are valid from $(A,B,C)$ and $(A,B',C)$ being local
structures in $G$ and $G'$ respectively.  In particular, every edge incident
to $B$ or $B'$ only has edges from $A$,  $C$, or $B, B'$.  And edges incoming to $A$
do not involved $B$.  These observations together allow to factor out all the terms
$p_{i, \pa(i)}(x_i| x_{\pa(i)})$ where $i \in V \setminus (B \cup C)$ or $i \in V' \setminus ( B' \cup C)$.

By the assumptions of being a local modification, we have that 
$V \setminus (B \cup C) =  V' \setminus (B' \cup C)$, and the edge structure of
edges with heads in $V \setminus (B \cup C) =  V' \setminus (B' \cup C)$ is exactly
the same.  Thus we have 
\[
\prod_{i \in V \setminus (B \cup C)} p_{i| \pa(i)}(x_i| x_{\pa(i)} )  = 
 \prod_{i \in V' \setminus (B' \cup C)} p_{i| \pa(i)}(x_i| x_{\pa(i)} ).
\]

On the other hand, we  know from above that 
\[
p_{G, C|A}(x_C| x_A) =  \sum_{x_B \in \calr_B} \prod_{i \in B \cup C} p_{i| \pa(i)}(x_i| x_{\pa(i)} )
\]
and
\[
p_{G', C|A}(x_C| x_A) = 
\sum_{x_{B'} \in \calr_{B'}} \prod_{i \in B' \cup C} p_{i| \pa(i)}(x_i| x_{\pa(i)} ).
\]

Since we have assumed that the two models for the two local structures produce 
exactly the same family of conditional probability distributions $p_{G, C|A}(x_C| x_A)$
and $p_{G', C|A}(x_C| x_A)$, this means
that the entire graphs $G_1$ and $G_2$ produce exactly the same probability distributions
when $X_B$ are hidden variables.
\end{proof}

Here we provide one common example of how  local modifications work.

\begin{prop}\label{prop:degree2vertex}
    Consider the displayed tree model on the DAG  $G = (V, D)$.  Let
    $a \rightarrow c$ be an edge.   Let $G' = (V', D')$  be the DAG obtained
    from $G$ by replacing $a \rightarrow c$ with the path $a \rightarrow b \rightarrow c$.
    If the transition model $TM$ is multiplicatively closed and splittable on 
    transition matrices, and $b$ is a hidden variable, then these two
    DAGs produce the same family of probability distributions on the
    observed variables.
\end{prop}

\begin{proof}
    First we show that $G$ and $G'$ are local modifications of each other.
    Then we show that under the phylogenetic model the appropriate families of conditional
    distributions are the same, applying Theorem \ref{thm:local}.

    To see that $G$ and $G'$ are local modifications, first we have to take
    appropriate local structures.  In $G$ we take $A = \pa(c)$, $B = \emptyset$,
    and $C = \{c\}$.  In $G'$ we have the same $A$ and $C$, and $B = \{b\}$.
    It is easy to check that these are both local structures.  Indeed, in $G$ nearly
    all conditions are vacuous since $B = \emptyset$, and all edges $i \rightarrow c$
    have $i \in A$ by construction.   Similarly, in $G'$, we have any edge $i \rightarrow b$
    is of the form $a \rightarrow b$, any edge $b \rightarrow i$ has $i = c$, and
    any edge $i \rightarrow c$ has $i = b$ or $i \in A$.  This shows that $G$ and $G'$
    are local modifications.

    Now we must show that we get the same families of conditional distributions
    $p_{c|A}(x_c| x_A)$ in both $G$ and $G'$.  Write $A = \{a\} \cup A'$.
    Then we have in $G$,
\[
p_{G, c|A}(x_c| x_A)  =  \pi^c_a M^{ac}(x_c|x_a)  + \sum_{a' \in A'} \pi^c_{a'} M^{a'c}(x_c|x_{a'})
\]
whereas in $G'$ we have 
\begin{eqnarray*}
    p_{G', c|A}(x_c| x_A) &  = & \pi^b_a \sum_{x_b} M^{ab}(x_b|x_a)M^{bc}(x_c|x_b)  + \sum_{a' \in A'} \pi^c_{a'} M^{a'c}(x_c|x_{a'}).
\end{eqnarray*}
We see that the distribution  $p_{G', c|A}(x_c| x_A)$ from $G'$ is a special case 
of the distribution $p_{G, c|A}(x_c| x_A)$ from $G$ by taking $\pi^c_a = \pi^b_a$,
and $M^{ac} = M^{ab} M^{bc}$, which is in the model since we assumed the class
of transition matrices in multiplicatively closed.

On the other hand, with a distribution from $G$, we reverse this process since
we assumed the model is splittable.    Indeed, we take $\pi^b_a = \pi^c_a$
and by the splittable property of our model, there exists an
$N$ in the model such that we can take $M^{bc} = N$ and $M^{ab} = M^{ac}N^{-1}$.
\end{proof}

As a second example, we show that we can contract the edge from a reticulation
vertex in some circumstances, without changing the set of probability distributions
that arise.

\begin{prop}\label{prop:contractretic}
     Consider the displayed tree model on a DAG $G$ which contains an edge $b \rightarrow c$.
     Suppose that the outdegree of $b$ is one
    and the indegree of $c$ is one.  Let $G'$ be the graph obtained from
    $G$ by contracting the edge $b \rightarrow c$ (that is, in $G'$, we
    have an edge $a \rightarrow c$ for each edge $a \rightarrow b$ in $G$).  
    If the transition model $TM$ is multiplicatively closed and 
    splittable, and $b$ is a hidden variable in $G$ then $G$ and $G'$ yield the same family of probability 
    distributions.
\end{prop}

\begin{proof}
    We need to show that $G$ and $G'$ are local modifications of each other,
    and that the corresponding conditional distributions families on 
    those local modifications are the same.

    In $G$, let $A = \pa(b)$, $B = \{b\}$, and $C = \{c\}$.  By construction
    $(A,B,C)$ is a local structure in $G$.  In $G'$, we denote the  vertex
    obtained by contracting the edge $b \rightarrow c$ by $c$ as well, and
    take $B' = \emptyset$.  Note that we then have edges $a \rightarrow c$ for 
    each $a \in A$.  Again, $(A, B', C)$ is a local structure in $G'$
    and the graphs are local modifications of each other.

    Now we need to prove that the family of conditional distributions $p_{G, c|A}(x_c|x_A)$ and $p_{G', c|A}(x_c|x_A)$ in the two
    graphs are the same.  In $G$ we have
\begin{eqnarray*}
   p_{G, c|A}(x_c | x_A) & = &  \sum_{x_b =1}^{r_b}  p_{b|A}(x_b| x_A) p_{c|b}(x_c|x_b)  \\
   & = &  \sum_{a \in A} \pi^b_a  \sum_{x_b=1}^{r_b} M^{ab}(x_b| x_a)  M^{bc}(x_c|x_b)  
\end{eqnarray*}
whereas in $G'$ we have
\begin{eqnarray*}
   p_{G', c|A}(x_c | x_A) & = &  \sum_{a \in A} \pi^c_a   M^{ac}(x_c| x_a)   \\
\end{eqnarray*}
Hence, we see that each conditional distribution from $G$ produces a distribution
from $G'$ by taking
$\pi^c_a =  \pi^b_a$ for all $a$, and $M^{ac}  =  M^{ab} M^{bc}$ for all $a$.  This
is valid because we assumed that the model is closed under multiplication of
transition matrices.

On the other hand, with a distribution from $G'$, we reverse this process since
we assumed the model is splittable.  Indeed, we take $\pi^b_a = \pi^c_a$
and by the splittable property of our model, there exists an
$N$ in the model such that we can take $M^{bc} = N$ and $M^{ab} = M^{ac}N^{-1}$ for each
$a$.  This shows that we can get exactly the same set of conditional distributions
from each graph, and hence  the two graphs give the same probability distributions
by Theorem \ref{thm:local}.
\end{proof}

\begin{figure}

\begin{tikzpicture}[scale=1, every node/.style={circle, draw, fill=white, inner sep=2pt}]
  \tikzset{edge/.style = {->,> = latex}}
    \node (1) at (-1,3.5) {1};
    \node (2) at (1,3.5) {2};
    \node (3) at (1,2.75)  {3};
    \node (4) at (-1,2) {4};
    \node (5) at (1,2) {5};
    \node (6) at (0,1) {6};
    \node (a) at (-2,4.5) {$\alpha$};
    \node (b) at (1,4.5) {$\beta$};
    \node (c) at (2,2.75) {$\gamma$};
    \node (d) at (0,0) {$\delta$};

    \draw[edge] (1) to (a);
    \draw[edge] (2) to (1);
    \draw[edge] (2) to (b);
    \draw[edge] (3) to (2);
    \draw[edge] (3) to (c);
    \draw[edge] (4) to (1);
    \draw[edge] (4) to (5);
    \draw[edge] (4) to (6);
    \draw[edge] (5) to (3);
    \draw[edge] (6) to (5);
    \draw[edge] (6) to (d);
    \end{tikzpicture}
\quad \quad
\begin{tikzpicture}[scale=1, every node/.style={circle, draw, fill=white, inner sep=2pt}]
  \tikzset{edge/.style = {->,> = latex}}
    \node (1) at (-1,3.5) {1};
    \node (2) at (1,3.5) {2};
    \node (3) at (1,2.75)  {3};
    \node (4) at (-1,2) {4};
   % \node (5) at (1,2) {5};
    \node (6) at (0,1) {6};
    \node (a) at (-2,4.5) {$\alpha$};
    \node (b) at (1,4.5) {$\beta$};
    \node (c) at (2,2.75) {$\gamma$};
    \node (d) at (0,0) {$\delta$};

    \draw[edge] (1) to (a);
    \draw[edge] (2) to (1);
    \draw[edge] (2) to (b);
    \draw[edge] (3) to (2);
    \draw[edge] (3) to (c);
    \draw[edge] (4) to (1);
    \draw[edge] (4) to (3);
    \draw[edge] (4) to (6);
    \draw[edge] (6) to (3);
    \draw[edge] (6) to (d);
    \end{tikzpicture}
    \quad \quad
\begin{tikzpicture}[scale=1, every node/.style={circle, draw, fill=white, inner sep=2pt}]
  \tikzset{edge/.style = {->,> = latex}}
    \node (1) at (-1,3.5) {1};
    \node (2) at (1,3.5) {2};
    \node (3) at (1,2.75)  {3};
    \node (4) at (-1,2) {4};
    \node (5) at (1,2) {5};
    \node (6) at (0,1) {6};
    \node (a) at (-2,4.5) {$\alpha$};
    \node (b) at (1,4.5) {$\beta$};
    \node (c) at (2,2) {$\gamma$};
    \node (d) at (0,0) {$\delta$};

    \draw[edge] (1) to (a);
    \draw[edge] (2) to (1);
    \draw[edge] (2) to (b);
    \draw[edge] (3) to (2);
    \draw[edge] (5) to (c);
    \draw[edge] (5) to (3);
    \draw[edge] (4) to (1);
    \draw[edge] (4) to (3);
    \draw[edge] (4) to (6);
    \draw[edge] (6) to (5);
    \draw[edge] (6) to (d);
    \end{tikzpicture}

\caption{Example of an unusual containment between displayed tree models. \label{fig:reticulationcontractex}}
\end{figure}

\begin{ex}
    Consider the three networks pictured in Figure \ref{fig:reticulationcontractex},
    where we assume that only the leaf vertices are observed.
    In the network on the left, the edge $5 \rightarrow 3$ satisfies the conditions of Proposition 
    \ref{prop:contractretic}.  So if the transition model is multiplicatively 
    closed and splittable,
    the network in the middle yields the same family of probability distributions
    as the network on the left.

    For the network on the right, note that the edge $5 \rightarrow 3$ does not
    satisfy the conditions of Proposition \ref{prop:contractretic}.  However,
    if our model of transition matrices contains the identity matrix, and
    we set this on that edge, then with that restriction, the edge does contract
    to produce the middle graph.  Thus, if our transition model is multiplicatively closed,
    splittable, and contains the identity matrix, the network model on the right
    contains all the distributions from the model on the left.  If we do not have the
    identity matrix as a transition matrix, but have the identity matrix as a limit,  then at this point, we can say that distributions in the
    model on the left appears as limits of distributions of the right.

    In the case of the Jukes-Cantor model, we checked the dimensions of the families of
    distributions from the left network and the right network and found that
    the dimension is 9 for the right network and 8 for the left network, so it
    is not possible that the two networks produce exactly the same family of distributions
    in that model.    It is unclear what happens for other transition matrix structures.    
\end{ex}

%%%%%%%%%%%%%%%%%%%%%%%%%%%%%%%%%%%%%%%%%%%%%
%%%%%%%%%%%%%%%%%%%%%%%%%%%%%%%%%%%%%%%%%%%%%
%%%%%%%%%%%%%%%%%%%%%%%%%%%%%%%%%%%%%%%%%%%%%
%%%%%%%%%%%%%%%%%%%%%%%%%%%%%%%%%%%%%%%%%%%%%

\section{Stacked Reticulations}  \label{sec:stacked}

In this section, we explore an extension  of 
Propositions \ref{prop:degree2vertex} and \ref{prop:contractretic}
which is concerned with stacked
reticulations.  A stacked reticulation is a pair of vertices
$b$, $c$ such that $b$ and $c$ are each reticulations (that is,
both have indegree greater than one), and such that $b \rightarrow c$
is an edge in the network.  We show that under fairly broad
conditions, it is possible to contract the edge $b \rightarrow c$
without changing the distributions that can arise from the
displayed tree model.  This result leads to 
a number of surprising cases where two binary networks
can yield the same probability distributions, and hence 
lead to non-identifiability of the network models.  
These results suggest that stacked reticulations should likely
be avoided in modeling phylogenetic networks under the displayed tree model.

\begin{thm}\label{thm:stacked}
    Consider the displayed tree model on a DAG $G$ which contains an edge $b \rightarrow c$. 
    Suppose that $b$ has outdegree $1$.
    Let $G'$ be the graph obtained from $G$ by contracting the edge $b \rightarrow c$
    (that is, we have an edge $a \rightarrow c$ in $G'$ for each edge $a \rightarrow b$
    in $G$, and we keep the graph simple).
    If the transition model $TM$ is multiplicatively closed, 
    closed under convex combinations, and
    splittable, then the two graphs produce the same family of distributions when $b$
    is a hidden variable in $G$.  The statement also holds if we require all
    reticulation probabilities to be positive.  
 \end{thm}

 \begin{proof}
    First, we show that the graphs $G$  and $G'$ are local modifications of each other.
    Then we show that the corresponding conditional distribution families are the same.

    For the graph $G$, take $B = \{b\}$, $C = \{c\}$ and $A = (\pa(b) \cup \pa(c)) \setminus \{b\}$.  Since we assumed there are no outgoing edges of $b$ besides
    $b \rightarrow c$, this shows that the triple $(A,B,C)$ is a local structure in 
    $G$.  In the graph $G'$ we take $B' = \emptyset$.  In $G'$ we just have the edges
    $a \rightarrow c$ for all $a \in A$.  So $(A, B', C)$ is a local structure
    on $G'$, and $G$ and $G'$ are local modifications of each other.

    Now we analyze the conditional distributions of $p_{G, c|A}$ and $p_{G', c|A}$
    in the two graphs.  In $G$ we use $M$ to denote the transition matrices, and $\delta$
    to denote reticulation parameters.  In $G'$, we use $N$ and $\epsilon$ respectively. To examine $p_{G, c|A}$ and $p_{G', c|A}$, we first split the set $A$ into three sets (based on $G$):
    $A_1$ is the set of vertices in $A$ have $b$ as a child but not $c$,
    $A_2$ is the set of vertices in $A$ that have $c$ as a child but not $b$.
     and
    $A_3$ is the set of vertices in $A$ that have both $b$ and $c$ as children.
    With this division of $A$, we have the following form for $p_{G, c|A}$:
\begin{eqnarray}\label{eq:Gevalstacked}
    p_{G, c|A}(x_c|x_A) & = &  \quad \sum_{a \in A_1} \delta^b_a \delta^c_b \sum_{x_b} M^{ab}(x_b|x_a) M^{bc}(x_c| x_b)    \\
    &   & +  \sum_{a \in A_2}  \delta^{c}_a  M^{ac}(x_c|x_a)  \nonumber \\
    &   &  + \sum_{a \in A_3}  \left(  \delta^b_a \delta^c_b \sum_{x_b} M^{ab}(x_b|x_a) M^{bc}(x_c| x_b) + \delta^{c}_a  M^{ac}(x_c|x_a)   \nonumber   \right).
\end{eqnarray}
On the other hand, in $G'$ we have
\begin{eqnarray*}
    p_{G', c|A}(x_c| x_A)  & = &  \sum_{a \in A} \epsilon^c_a N^{ac}(x_c | x_a).
\end{eqnarray*}

It is straightforward to see how to use the parameters from $G$ to produce 
parameters in $G'$.  Indeed, for
$a \in A_1$, take 
\[
\epsilon^c_a =  \delta^b_a \delta^c_b,  \quad  N^{ac} =  M^{ab}M^{bc}.
\]    
For $a \in A_2$ take
\[
\epsilon^c_a =  \delta^c_a  \quad  N^{ac} = M^{ac}.
\]
And for $a \in A_3$ take
\[
\epsilon^c_a  =  \delta^b_a \delta^c_b + \delta^c_a  \quad  N^{ac}  =  \frac{\delta^b_a \delta^c_b}{\delta^b_a \delta^c_b + \delta^c_a}  M^{ab}M^{bc}  
+  \frac{\delta^c_a}{\delta^b_a \delta^c_b + \delta^c_a}  M^{ac} 
\]
Since the substitution model $TM$ is closed under matrix multiplication and convex
combinations, this produces reticulation probabilities and transition matrices
that belong to the conditional distribution for $c$ given $A$ in $G'$
and if all the $\delta$'s are positive so are the epsilons.

Now we need to show that any conditional distribution $p_{G', c|A}$ can be obtained
from parameters from $G$.  Note that the parametrization from $G$ has many more
free parameters than $G'$, so there might be many ways to do this.

First of all, for $a \in A_2$, we take
\[
\delta^c_a = \epsilon^c_a     \quad  M^{ac} = N^{ac}.
\]
Let $M^{bc}$ be a transition matrix
in our model so that for each $a \in A_1 \cup A_3$
\[
M^{ab} = N^{ac}(M^{bc})^{-1}
\]
is a transition matrix in the model, for each $a$.  The matrix $M^{bc}$
exists because the model is splittable.  For $a \in A_3$ take
\[
M^{ac}  =  N^{ac}.
\]
We set
\[
\delta^{c}_b = \sum_{a \in A_1} \epsilon^c_a  +  \frac{1}{2} \sum_{a \in A_3} \epsilon^c_a .
\]
For $a \in A_1$, we take
\[
\delta^b_a =  \frac{\epsilon^c_a}{\delta^{c}_b}.
\]
For $a \in A_3$, we take
\[
\delta^b_a =  \frac{\epsilon^c_a}{2 \delta^{c}_b} \quad \quad \delta^c_a =
\frac{1}{2} \epsilon^c_a.
\]
It is straightforward to see that these choices for the parameters in $G$
will yield the parameters in $G'$ when plugged into the formulas of Equation
\ref{eq:Gevalstacked}.  The most complicated to check are for $a \in A_3$,
but we see that
\[
 \frac{\delta^b_a \delta^c_b}{\delta^b_a \delta^c_b + \delta^c_a}  M^{ab}M^{bc}  
+  \frac{\delta^c_a}{\delta^b_a \delta^c_b + \delta^c_a}  M^{ac} = \frac{1}{2} N^{ac} + \frac{1}{2} N^{ac}  = N^{ac}
\]
\[
\delta^b_a \delta^c_b + \delta^c_a  = \frac{1}{2} \epsilon^{c}_{a} + 
\frac{1}{2} \epsilon^{c}_{a} =  \epsilon^{c}_{a} 
\]
as desired.  Furthermore, if all of $\epsilon$'s were positive, then the
constructed $\delta$'s will be also.
\end{proof}

\begin{rmk}
    Note that in terms of the conditions on the network, Theorem \ref{thm:stacked} appears to be
    a more general result than Propositions \ref{prop:degree2vertex} and \ref{prop:contractretic}.  However,
    Theorem \ref{thm:stacked} requires a more restrictive assumption on the transition model $TM$ (specifically,
    it requires $TM$ to be closed under convex combinations).   So Propositions  \ref{prop:degree2vertex} and \ref{prop:contractretic}
    are not a special case of Theorem \ref{thm:stacked}.
\end{rmk}

\begin{figure}

  \begin{tikzpicture}[scale=1, every node/.style={circle, draw, fill=white, inner sep=2pt}]
  \tikzset{edge/.style = {->,> = latex}}
    \node (1) at (2,3) {1};
    \node (2) at (1,2.5) {2};
    \node (3) at (2,2) {3};
    \node (4) at (1,1.5) {4};
    \node (5) at (2,1) {5};
    \node (a) at (0,2) {$\alpha$};
    \node (b) at (2,0) {$\beta$};
    \node (c) at (3,2.5) {$\gamma$};
    
    \draw[edge] (1) to (2);
     \draw[edge] (1) to (3);
    \draw[edge] (2) to (4);
    \draw[edge] (3) to (4);
     \draw[edge] (3) to (5);
     \draw[edge] (4) to (5);
     \draw[edge] (2) to (a);
     \draw[edge] (1) to (c);
     \draw[edge] (5) to (b);
  \end{tikzpicture}  \quad \quad
   \begin{tikzpicture}[scale=1, every node/.style={circle, draw, fill=white, inner sep=2pt}]
  \tikzset{edge/.style = {->,> = latex}}
    \node (1) at (2,3) {1};
    \node (2) at (1,2.5) {2};
    \node (3) at (2,2) {3};

    \node (5) at (2,1) {5};
    \node (a) at (0,2) {$\alpha$};
    \node (b) at (2,0) {$\beta$};
    \node (c) at (3,2.5) {$\gamma$};
    
    \draw[edge] (1) to (2);
     \draw[edge] (1) to (3);
    \draw[edge] (2) to (5);
     \draw[edge] (3) to (5);
     \draw[edge] (2) to (a);
     \draw[edge] (1) to (c);
     \draw[edge] (5) to (b);
  \end{tikzpicture}
  \quad \quad
   \begin{tikzpicture}[scale=1, every node/.style={circle, draw, fill=white, inner sep=2pt}]
  \tikzset{edge/.style = {->,> = latex}}
    \node (1) at (2,3) {1};
    \node (2) at (1,2.5) {2};
    \node (5) at (2,1) {5};
    \node (a) at (0,2) {$\alpha$};
    \node (b) at (2,0) {$\beta$};
    \node (c) at (3,2.5) {$\gamma$};
    
    \draw[edge] (1) to (2);
     \draw[edge] (1) to (5);
    \draw[edge] (2) to (5);
     \draw[edge] (2) to (a);
     \draw[edge] (1) to (c);
     \draw[edge] (5) to (b);
  \end{tikzpicture}

  \caption{\label{fig:stackedretic} A stacked reticulation contraction of the edge
  $4 \rightarrow 5$.   All three networks
  produce the same probability distributions on $\alpha, \beta, \gamma$.}

\end{figure}
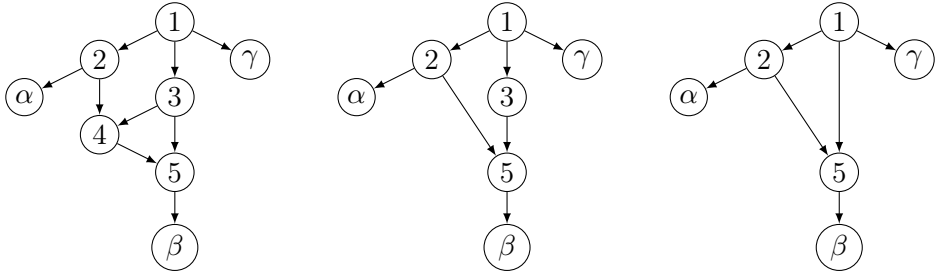

\begin{ex}
   Consider the graphs in Figure \ref{fig:stackedretic}.  The network on the left
   has a stacked reticulation, and the local structure with $A = \{2,3\}, B = \{4\}, C = 
   \{5\}$, and satisfies the conditions of Theorem \ref{thm:stacked}.
   Hence, this graph yields the same probability distributions on the observed leaves
   $\alpha, \beta, \gamma$ as the network in the middle.  
   Proposition \ref{prop:degree2vertex} means that the network in the middle
   yields the same probability distributions as the network on the right.
   So this is a situation where a level-$2$ binary network
   will produce exactly the same family of probability distributions as a 
   level-$1$ binary network.    
\end{ex}

\begin{ex}
    Consider the three networks in Figure \ref{fig:stacked2}.  Comparing the networks
    on the outside with the networks in the middle, we see that we can apply
    Theorem \ref{thm:stacked} to see that they yield the same families of probability
    distributions on observed leaves $\alpha, \beta, \gamma,  \delta$.  In particular,
    in both networks the stacked reticulation is on the edge $3 \rightarrow 5$.
    Contracting that edge in both cases yields the network in the middle.
\end{ex}

One application of the stacked reticulation results concerns the presence of
2-blobs inside of a network, and the fact that these cannot be
identified under the displayed trees model.   Non-identifiability of $2$-cycles in the
displayed tree model is a common observation when studying the equivariant models \cite{GrossLong2018, Grossvan2021},
and Propositions \ref{prop:2blob} and \ref{prop:rootblob} are a generalization of those results for arbitrary $2$-blobs.

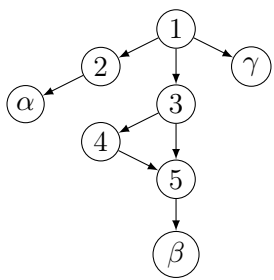
\begin{figure}
\begin{center}
 \begin{tikzpicture}[scale=1, every node/.style={circle, draw, fill=white, inner sep=2pt}]
  \tikzset{edge/.style = {->,> = latex}}
    \node (1) at (2,3) {1};
    \node (2) at (1,2.5) {2};
    \node (3) at (2,2) {3};
    \node (4) at (1,1.5) {4};
    \node (5) at (2,1) {5};
    \node (a) at (0,2) {$\alpha$};
    \node (b) at (2,0) {$\beta$};
    \node (c) at (3,2.5) {$\gamma$};
    
    \draw[edge] (1) to (2);
     \draw[edge] (1) to (3);
    \draw[edge] (3) to (4);
     \draw[edge] (3) to (5);
     \draw[edge] (4) to (5);
     \draw[edge] (2) to (a);
     \draw[edge] (1) to (c);
     \draw[edge] (5) to (b);
  \end{tikzpicture}
\caption{A network with a $2$-blob consisting of vertices $\{3,4,5\}$.}  \label{fig:2blob}
\end{center}

\end{figure}

Recall that a \emph{blob}
in a network is a maximal $2$-connected subgraph (that is, a subgraph that
cannot be disconnected by deleting an edge).  A $2$-blob is a blob in
a network that has two edges that connect it to other parts of the network.
An example of a $2$-blob appears in Figure \ref{fig:2blob}, with the $2$-blob itself being
the subgraph consisting on vertices $\{3,4,5 \}$.

We distinguish two types of $2$-blobs, depending on whether the root is part of the $2$-blob, or not.
If the root is part of the $2$-blob we call it a \emph{root blob}, otherwise it is a \emph{non-root blob}.
If a $2$-blob is a root blob, there are two vertices $b_1$ and $b_2$ that are sinks of the blob,
with edges $b_1 \rightarrow c_1$  and $b_2 \rightarrow c_2$ out of the blob.  
If a $2$-blob is a non-root blob, the blob has a source $b_1$ and a sink $b_2$, with an edge $a \rightarrow b_1$
and an edge $b_2 \rightarrow c$, into and out of the blob, respectively.

\begin{prop}\label{prop:2blob}
Let $G$ be a DAG with a non-root $2$-blob, and let $B$ be the set of vertices in the $2$-blob.
Let $a \rightarrow b_1$ be the edge pointing into the $2$-blob and 
$b_2 \rightarrow c$ be the edge pointing out of the $2$-blob.  Suppose that
all the random variables $X_b$ with $b \in B$ are hidden variables.  Let $G'$
be the graph obtained from $G$ by deleting all the vertices in $B$ and incident edges
and adding the edge $a \rightarrow c$.
If the transition model $TM$ is multiplicatively closed, closed under convex combinations, and
    splittable, then the two graphs $G$ and $G'$ produce the same probability distributions.
\end{prop}

\begin{proof}
   We use Theorem \ref{thm:stacked} 
   repeatedly.  First of all, because we have assumed that $TM$
   is closed under convex combinations, we can assume that the DAG $G$ has no
   parallel edges/$2$-cycles.  Indeed, if there were $k$  parallel edges $e \to f$,
   in the displayed tree model, the conditions distribution of $X_f| X_e$ would be
   \[
   \sum_{i = 1}^k\pi^{f,i}_e M^{ef,i}
   \]
   where the $i$ in the superscripts is the indicator for the $i$th of the $k$ copies of the edge $e \to f$.  
   
   We can assume that all degree $2$ vertices within the $2$-blob have been
   \seth{suppressed} by using Proposition \ref{prop:degree2vertex}.  
   Then, since a $2$-blob $B$ does not consist only of the single edge $b_1 \rightarrow b_2$, then
   $b_2$ must have $2$ or more parent vertices.  None of those vertices can have a child
   that is outside of $B$.  Since $b_2$ is a sink of $B$, there must be a vertex $b_3$
   in $B$ whose outdegree is $1$, with an edge $b_3 \rightarrow b_2$.  By
   Theorem \ref{thm:stacked}, we can contract this edge without changing the family
   of distributions that can arise.  This process can be repeated until we are left with only
   the edge $b_1 \rightarrow b_2$. Both of the degree two vertices $b_1$ and $b_2$
   can be contracted using Proposition \ref{prop:degree2vertex}.  The end result
   of this process is the graph $G'$.
 \end{proof}

In the case of a root $2$-blob, we end up with a weaker statement, because not every root $2$-blob can be eliminated using the removal of stacked reticulations.
We also need the concept of closed under marginal convex combinations (Definition
\ref{defn:closedmarginal}), \seth{and a common condition of a phylogenetic
model we call \emph{root unidentifiable}}.

\seth{
\begin{defn}
    A pair of transition matrices $TM$ and root distributions $RD$ is \emph{root unidentifiable} if for any $M^1 \in TM$ and $\pi^1 \in RD$, there are
    $M^2 \in TM$ and $\pi^2 \in RD$ such that
    \[
    M^1(x_b| x_a) \pi^1(x_a)  =  M^2(x_a | x_b) \pi^2(x_b)  
    \]
    for all $x_a, x_b \in \Sigma$.
\end{defn}

The name ``root unidentifiable'' is used because this condition means 
that any joint distribution from the model obtained from the graph $1 \rightarrow 2$,
could also have been obtained from the graph $2 \rightarrow 1$, with a different choice
of rate matrix and root distribution.  This property occurs in  many models (for example,
all equivariant models satisfy it) and it implies that in a larger tree, the location
of the root cannot be identified since it is possible to switch directions of the
edges without changing the resulting probability distributions that arise.  
}

 \begin{prop}\label{prop:rootblob}
 Let $G$ be a DAG with a root $2$-blob, and let $B$ be the set of vertices in the $2$-blob.
 Let $b_1$ and $b_2$ be the two sinks of the blob, with outgoing edges $b_1 \rightarrow c_1$
 and $b_2 \rightarrow c_2$.  Suppose all random variables $X_b$ with $b \in B$ are hidden variables.  
 Let $G'$
be the graph obtained from $G$ by deleting all the vertices in $B$ except $b_1$ and $b_2$, and adding
a vertex $a$ with edges $a \rightarrow b_1$ and $a \rightarrow b_2$, and having $X_a$ be a hidden variable.
If the transition model $TM$ is multiplicatively closed, 
closed under marginal convex combinations, and 
\seth{root unidentifiable}, then every distribution from $G$ can be realized as a distribution from
$G'$.  
  \end{prop}

\begin{proof}
    It suffices to compute the marginal distribution of $(X_{b_1}, X_{b_2})$ in both graphs.
    To see this, let $C$ be the set of all vertices below the $2$-blob.
       Since the  $2$-blob is ancestral to all other vertices in $G$,
       so the marginal distribution on $C' = \{b_1, b_2\} \cup C$ is
\begin{eqnarray*}
  p_{G,C }(x_{b_1}, x_{b_2}, x_C)  & = &  p_{G,\{b_1, b_2\}  }(x_{b_1}, x_{b_2}) p_{G,C|\{b_1, b_2\}  }(x_C | x_{b_1}, x_{b_2})    \\
   p_{G',C }(x_{b_1}, x_{b_2}, x_C)  & = &  p_{G',\{b_1, b_2\}  }(x_{b_1}, x_{b_2}) p_{G',C|\{b_1, b_2\}  }(x_C | x_{b_1}, x_{b_2}). 
\end{eqnarray*}
    However, 
    \[
    p_{G,C|\{b_1, b_2\}  }(x_C | x_{b_1}, x_{b_2})   = p_{G',C|\{b_1, b_2\}  }(x_C | x_{b_1}, x_{b_2}) 
    \]
    by assumption since that only involves edges that are not in the blob.  
         Hence, we can suppose that $G$ just consists of the $2$-blob itself,
    and $G'$ is the graph with only three vertices $a, b_1, b_2$, and edges $a \rightarrow b_1$,
    $a \rightarrow b_2$.

    In the graph $G$, the marginal distribution is obtained by summing over all possible
    ways to choose reticulation edges at each reticulation vertex.  The resulting probability
    distribution when each of the those reticulation edge choices are made is a $2$-leaf tree.
    Those distributions for $2$-leaf trees are weighted by the reticulation probabilities
    associated to each of those edge choices.  Since the model is \seth{root unidentifiable} 
    and multiplicatively closed,  each one of those distributions is 
    a marginal distribution from the model on a two leaf tree, which belongs
    to the associated set of 2-variable marginal distributions, $MD$. 
    Since the transition model is closed under marginal convex combinations, this shows that
    every distribution from $G$ can be realized as a distribution from $G'$.
\end{proof}

On the other hand, every distribution coming from the simplified graph $G'$ arises
as a limit of distributions from $G.$  
This gives a partial converse to Proposition \ref{prop:rootblob}.

 \begin{prop}\label{prop:limitrootblob}
 Let $G$ be a DAG with a root $2$-blob, and let $B$ be the set of vertices in the $2$-blob.
 Let $b_1$ and $b_2$ be the two sinks of the blob, with outgoing edges $b_1 \rightarrow c_1$
 and $b_2 \rightarrow c_2$.  Suppose all random variables $X_b$ with $b \in B$ are hidden variables.  
 Let $G'$
be the graph obtained from $G$ by deleting all the vertices in $B$ except $b_1$ and $b_2$, and adding
a vertex $a$ with edges $a \rightarrow b_1$ and $a \rightarrow b_2$, and having $X_a$ be a hidden variable.

Suppose that $TM$ is splittable and contains transition matrices that are arbitrarily close
to the identity matrix.  Then any distribution produced by the network $G'$ is in the limit of the set of distributions produced
by the network $G$.
\end{prop}

\begin{proof}
    As in the proof of Proposition \ref{prop:rootblob}, we can suppose that $G$ just consists of the $2$-blob itself,
    and $G'$ is the graph with only three vertices $a, b_1, b_2$, and edges $a \rightarrow b_1$,
    $a \rightarrow b_2$.
So a distribution $P$ that is a marginal distribution associated to the graph $G'$ is of
    the form 
    \[
    P(x_{b_1}, x_{b_2}) = \sum_{x_a}  \pi(x_a) M^{ab_1}(x_{b_1} | x_a) M^{ab_2}(x_{b_2} | x_a).
    \]

    In $G$, let $p_1$ be any path from the root to $b_1$ and let $p_2$ be any path from the root
    to $b_2$.  Let the vertices on $p_1$ be $a, c_1, \ldots, c_k, b_1$
    and the vertices on the path $p_2$ be $a, d_1, \ldots, d_l, b_2.$   
    
    Suppose that $p_1$ and $p_2$ do not have any common vertices besides the root $a$.
    Since $TM$ is splittable, we can choose a sequence of matrices, 
    $M^{ac_1}, M^{c_1c_2}, \ldots, M^{c_kb_1} \in TM$ such that 
    \[
M^{ab_1}  = M^{ac_1}M^{c_1c_2} \cdots M^{c_kb_1}.
    \]
    Similarly, there is a sequence of matrices $M^{ad_1}, M^{d_1d_2}, \ldots, M^{d_lb_2} \in TM$
    such that 
    \[
    M^{ab_2}  = M^{ad_1}M^{d_1d_2} \cdots M^{d_lb_2}.
    \]
    For each of the edges in the paths $p_1$ and $p_2$  assign the corresponding transition
    matrix produced above.  Make the root distribution $\pi$ in $G$ the same as the root
    distribution in $G'$.  For all the other edges in $G$, take arbitrary matrices in $TM$.
    For the reticulation parameters, for each reticulation that involves one of the edges
    on either the path $p_1$ or $p_2$, choose the probability $1-\epsilon$ for the edge
    involved in the path, and distribute $\epsilon$ among the other reticulation edges (since
    there may be more than two edges entering a reticulation).  For all other
    reticulations, choose the reticulation probabilities arbitrarily.

    When $\epsilon = 0$, the marginal probability of $(X_{b_1}, X_{b_2})$ is completely
    determined by the probability on the pair of paths $p_1$ and $p_2$, which is equal
    to the distribution $P$.  Hence, as $\epsilon \rightarrow 0$, the probabilities
    in the model for $G'$ converge to the distribution $P$.

    Finally, suppose that $p_1$ and $p_2$  share a vertex besides the root.  In that case,
    we can treat the lowest shared such vertex as the root, and use the argument above.
    Indeed, in addition to the argument above, we use, for each edge not on the paths
    $p_1$ and $p_2$, a sequence of matrices $M_\epsilon$ converging to the identity matrix as
    $\epsilon \rightarrow 0$ and put that matrix $M_\epsilon$ on each edge not in $p_1$ and $p_2$.  
    \end{proof}

Note that in Proposition \ref{prop:limitrootblob}, we needed to take a limit, because
it is typically assumed that reticulation probabilities should all be in $(0,1)$ on
a network.  Similarly, a more simple approach could also involve using identity matrices
on most edges, but that might be excluded by our considerations on the transition model $TM$,
e.g.~if we want to assume all branch lengths are $> 0$.

The combination of Propositions \ref{prop:rootblob} and \ref{prop:limitrootblob}
is usually enough that one can ignore root $2$-blobs in identifiability analysis of 
the displayed tree model.  Indeed, if we consider a model that is a semialgebraic
set for each network $G$, then if $G$ and $G'$ are related as above, and 
the transition model is multiplicatively closed, 
closed under marginal convex combinations,  
time reversible, and splittable, then the two models $M_G$ and $M_{G'}$ have the
same topological closure, and $M_G \subseteq M_{G'}$.  Hence, if $H$ is another
network and $M_{G'}$ and $M_H$ are distinguishable, then so are $M_G$ and  $M_{H}$.

It remains an open problem to show that in the case of a root $2$-blob, the two
networks $G$ and $G'$ produce the exact same probability distributions, under a suitable
restriction on the transition model $TM$.

%%%%%%%%%%%%%%%%%%%%%%%%%%%%%%%%%
%%%%%%%%%%%%%%%%%%%%%%%%%%%%%%%%%
%%%%%%%%%%%%%%%%%%%%%%%%%%%%%%%%%
%%%%%%%%%%%%%%%%%%%%%%%%%%%%%%%%%

\section{Non-Identifiability of Numerical Parameters through Conditional Distributions}
\label{sec:nonidentcond}

In this section, we point out some loss of identifiability that can occur
in the displayed tree model.
In particular, it can happen that the conditional distribution
at a reticulation node has lower dimension than the number of parameters
that go into it, which results in a loss of identifiability in those numerical 
parameters.  

To this end, consider the family of conditional distributions
in the displayed tree model at a single reticulation node.
Recall that a general conditional distribution at a reticulation
node will have the form
\[
p_{c | \pa(c)}(x_c | x_{\pa(c)})  =  \sum_{a \in \pa(c)} \pi^c_a  M^{ac}(x_c|x_a)
\]
which gives a restricted class of conditional distributions.
In fact, we can see that if $m = \#\pa(c)$ and we consider random variables
with $k$ states, then in the completely general DAG model, the dimension
of the space of such conditional distributions is $(k-1)k^m$.  This is because
for each of the $k^m$ states of $x_{\pa(c)}$ we get a probability distribution in $\Delta_k$, which has $k-1$ free parameters.

On the other hand, in terms of parametrizing the set of conditional distributions
that arise from the network model, we see that there are $m-1$ parameters for the $\pi^c_a$
parameters, and each transition matrix gives $(k-1)k$ parameters, 
for a total of $m(k-1)k + m-1$ parameters.  There are even fewer parameters
for other equivariant models.

It turns out that the conditional distributions that arise from network model
have to satisfy many linear relations.  We will focus just on the case of the
general Markov model for an arbitrary reticulation.  

\begin{prop}\label{prop:lineargeneral}
    Consider a general reticulation vertex $c$ with $\pa(c) = A$, and edges $a \rightarrow c$ for $a \in A$
    as part of the displayed tree model.
    Then the conditional distribution $p_{c|A}(x_c|x_A)$ satisfies
    the relations
    \[
p_{c|A}(x_c|x_A) + p_{c|A}(x_c|y_A)  = 
    p_{c|A}(x_c|x'_A) + p_{c|A}(x_c|y'_A) 
    \]
    for all $x_c \in \Sigma$, $x_A, y_A \in \Sigma^A$,  where
    $x'_A$ and $y'_A$ are any vectors of states such that for all $a \in A$
    $\{x_a, y_a \}  =  \{x'_a, y'_a \}$ as sets.  
\end{prop}

\begin{proof}
    For a distribution in the network model we have
    \begin{eqnarray*}
       p_{c|A}(x_c|x_A) + p_{c|A}(x_c|y_A) & = & \sum_{a \in A} \pi^c_a  M^{ac}(x_c|x_a) + \sum_{a \in A}  \pi^c_a  M^{ac}(x_c|y_a)\\
       & = &  \sum_{a \in A} \pi^c_a(M^{ac}(x_c| x_a) + M^{ac}(x_c | y_a) ) \\
       & = &  \sum_{a \in A} \pi^c_a(M^{ac}(x_c| x'_a) + M^{ac}(x_c | y'_a) ) \\
        & = & \sum_{a \in A} \pi^c_a  M^{ac}(x_c|x'_a) + \sum_{a \in A} \pi^c_a   M^{ac}(x_c|y'_a)\\
       & = & p_{c|A}(x_c|x'_A) + p_{c|A}(x_c|y'_A).
    \end{eqnarray*}
    This follows because the condition  that for all $a \in A$ we have 
    $\{x_a, y_a \}  =  \{x'_a, y'_a \}$ guarantees that the same terms appear in both 
    sums for each $a$.
\end{proof}

The equations from Proposition \ref{prop:lineargeneral} restrict the conditional
distributions that come from the displayed tree model to a low dimensional space.

% Furthermore, if the model is an equivariant model, then there might be further
% restrictions that come on the transition matrices that come from the equivarient
% conditions.

% \begin{prop}\label{prop:linearequiv}
% Consider an equivariant model with group $G$, and state space $\Sigma$.  Consider the three vertex subgraph with edges $a \rightarrow c$, $b \rightarrow c$
%     as part of a phylogenetic network model.  Then the conditional 
%     distribution $p_{c|\{a,b\}}(x_c|x_a, x_b)$ satisfies:
%     \[
%     p_{c|\{a,b\}}(x_c|x_a, x_b)  =  p_{c|\{a,b\}}(g(x_c)|g(x_a), g(x_b))
%     \]
% for all $x_a, x_b, x_c \in \Sigma$ and $g \in G$.
% \end{prop}

% \begin{proof}   
% Since the model is equivariant we have that
% \[
% M^{e}(g(x_j) | g(x_i))  =  M^e(x_j | x_i)
% \]
% for all transition matrices $M^e$, all $x_i, x_j \in \Sigma$ and all $g \in G$.
% Thus
% \begin{eqnarray*}
%     p_{c|\{a,b\}}(g(x_c)|g(x_a), g(x_b))  & = &   \pi^c_a  M^{ac}(g(x_c)|g(x_a)) + \pi^c_b  M^{bc}(g(x_c)|g(x_b) ) \\
%     & = &   \pi^c_a  M^{ac}(x_c|x_a) + \pi^c_b  M^{bc}(x_c|x_b)  \\
%     & = &  p_{c|\{a,b\}}(x_c|x_a, x_b).  
% \end{eqnarray*}
% \end{proof}

We now consider consequences for different phylogenetic modeling situations.

\begin{thm}\label{thm:dimloss}
    Consider the general Markov model on $k$ states, and a reticulation
    node $c$ with $\pa(c) = A$.  Let $m = \#A$.  Then the space of
    conditional distributions $p_{c|A}$ that can arise from this model
    has dimension $(k-1)(m(k-1) + 1)$.  In particular, since
    there are there $m(k-1)k + m-1$ parameters that go into the displayed trees
    model to describe this conditional distribution, this results in 
    a loss of $(m - 1)k$ dimensions.
\end{thm}

\begin{proof}
    We can use  Proposition \ref{prop:lineargeneral}
    to calculate the number of linearly independent parameters
    in the linear space of conditional distributions $p_{c|A}$ that could come
    from the displayed tree model.
    Indeed, for each fixed value of $x_c$ consider the set of coordinates
\[    
    \{p(x_c| x_A):   \#\{ a \in A : x_a \neq 1 \}  \leq 1 \} .  
\]    
  If we add any coordinate $p(x_c|x_A)$ with $x_A$ outside of this set, there
  will be a linear relation using the relations from Proposition \ref{prop:lineargeneral}.

On the other hand, we claim that the set of coordinates
\[
S = \{p(x_c| x_A):  x_c \in [k-1] \mbox{ and } \#\{ a \in A : x_a \neq 1 \}  \leq 1  \}
\]
is algebraically independent.  Indeed, for each fixed $x_c$, there
is no overlap of the entries of the matrices $M^{ac}$ that are involved
(and we leave off $x_c = k$, since otherwise we would get a relation since
$\sum_{x_c} p_{c|A}(x_c | x_A) = 1$).  
On the other hand, for each $x_A$, where $x_a \neq 1$ and each $x_c$,
$p_{c|A}(x_c|x_A)$ is the unique place where the algebraically independent
entry $M^{ac}(x_c|x_a)$ appears.  Thus shows that $S$ is algebraically independent.

Note that $\#S =  (k-1)(m(k-1) + 1)$.  On the other hand, the number of
free parameters in each matrix is $k(k-1)$, there are $m$ of them, and there are
$m-1$ reticulation parameters for a total of $mk(k-1) + m-1$.  But
\[
mk(k-1) + m-1  - ( (k-1)(m(k-1) + 1))  =  (m-1)k
\]
which gives the indicated loss of dimension. 
\end{proof}

Generalizations of Theorem \ref{thm:dimloss} for
general equivariant models appear in \cite{Casanellas2025}, including
a useful reparametrization of the conditional distribution when the
reticulation node only has two parents.

%%%%%%%%%%%%%%%%%%%%%%%%%%%%%%%%%
%%%%%%%%%%%%%%%%%%%%%%%%%%%%%%%%%
%%%%%%%%%%%%%%%%%%%%%%%%%%%%%%%%%
%%%%%%%%%%%%%%%%%%%%%%%%%%%%%%%%%

\section{Conditional Independence and the Ranks of Flattenings}  \label{sec:rankofflat}

In this section, we explore how conditional independence structures
in graphical models can be used to deduce new results about rank conditions
on flattening matrices in probability distributions that come from the 
displayed tree model.  These rank of flattening results generalize the classic
results for trees \cite{Allman2008}.  For level-1 networks, ranks for flattenings
can be used as a tool to prove identifiability results in the displayed tree model \cite{Casanellas2025}.

First, we recall the definition of conditional independence
of random vectors.

\begin{defn}
    Let $X$ be a discrete random vector with index set $V$ and let $A$, $B$, and $C$
    be disjoint subsets of $V$.  We say that $X_A$ is conditionally independent
    of $X_B$ given $X_C$ (denoted $X_A \ind X_B | X_C$) if for all $x_A \in \calr_A$,
    $x_B\in \calr_B$, and $x_C \in \calr_C$ we have
    \[
    P(X_A = x_A, X_B = x_B | X_C = x_C)  =  P(X_A = x_A | X_C = x_C) P(X_B = x_B | X_C = x_C).
    \]
\end{defn}

The key concept to describe conditional independence constraints
in DAGs 
is the notion of $d$-separation.

\begin{defn}
Let $G = (V,D)$ be a DAG.
    \begin{itemize}
        \item A \emph{chain} in a digraph is a sequence of vertices
        $\pi = v_1, v_2, \ldots, v_k$ such that for each $i$ either
        $v_i \rightarrow v_{i+1}$ or $v_{i+1} \rightarrow v_i$
        is in $D.$
        \item  A \emph{collider} is a sequence of vertices $a, b, c$
        such that $a \rightarrow b$ and $c \rightarrow b$ are edges.
        \item  A chain $\pi$ from $a$ to $b$ is said to be \emph{blocked}
        by a set $S \subseteq V$ if there is a vertex $v_i \in \pi$
        such that either \begin{itemize}
            \item $v_i \in S$ and $v_{i-1}, v_i, v_{i+1}$ is not a collider or
            \item $v_i$ and all its descendants are not in $S$
            and $v_{i-1}, v_i, v_{i+1}$ is a collider.
            \end{itemize}  
        \item  Two sets of vertices $A$ and $B$ are said to be $d$-separated given
        a set $C$ if for all $a \in A$ and $b \in B$, every chain from
        $a$ to $b$ is blocked by $C$.
    \end{itemize}
\end{defn}

\begin{prop}\label{prop:dsep}
    A conditional independence statement $X_A \ind X_B | X_C$ holds for
    all distributions associated to the DAG $G$ if and only if
    $A$ and $B$ are $d$-separated by $C$ in $G$.
 \end{prop}

See \cite{Lauritzen1996} Section 3.2.2 for more on d-separation and a proof
of Proposition \ref{prop:dsep}.

The conditional independence statement being true for a probability distribution
 means that certain rank conditions hold
 on flattened and marginalized versions of the joint probability distribution.

\begin{defn}
    Let $A, B, C$ be disjoint subsets of $V$.  Let $P$ be a joint distribution.
    The \emph{flattening matrix}
    ${\rm Flat}(A,B)(P)$ is the matrix whose rows are indexed by $\calr_A$ and columns are indexed by
 $\calr_B$, and where the entry in the $(x_A, x_B)$ row and column is the marginal
 probability
 \[
 P(X_A = x_A, X_B = x_B).
 \] 
       The \emph{conditional flattening
    matrix} is the matrix
 $   {\rm Flat}(A,B | C, x_C)(P) $
 is the matrix whose rows are indexed by $\calr_A$ and columns are indexed by
 $\calr_B$, and where the entry in the $(x_A, x_B)$ row and column is the probability
 \[
 P(X_A = x_A, X_B = x_B,  X_C = x_c).
 \]    
\end{defn}    
 
\begin{ex}
Suppose that we have 5 random variables, $X_1, \ldots, X_5$ each of which is binary.  Let
$A = \{1,2\}, B = \{3,4\}, C = \{5\}$.  Let
\[
p_{x_1x_2x_3x_4x_5}  =  P(X_1 = x_1, X_2 = x_2, X_3 = X_3, X_4 = x_4, X_5 = x_5).
\]
Then
\[
{\rm Flat}(12,34 | 5, x_5)(P) = 
\begin{pmatrix}
    p_{0000x_5} & p_{0001x_5} & p_{0010x_5} & p_{0011x_5} \\
    p_{0100x_5} & p_{0101x_5} & p_{0110x_5} & p_{0111x_5} \\
    p_{1000x_5} & p_{1001x_5} & p_{1010x_5} & p_{1011x_5} \\
    p_{1100x_5} & p_{1101x_5} & p_{1110x_5} & p_{1111x_5} \\
\end{pmatrix}
\]    
\end{ex}

\begin{prop}
    Let $P$ be a probability distribution that satisfies the conditional
    independence statement $X_A \ind X_B | X_C$.  Then for each $x_C \in \calr_C$,
    the conditional flattening matrix $   {\rm Flat}(A,B | C, x_C)(P) $
    has rank $\leq 1$.
\end{prop}

\begin{proof}
    This follows from the definition of conditional independence given above.
    Indeed, we have the equation
    \[
    P(X_A = x_A, X_B = x_B | X_C = x_C)  =  P(X_A = x_A | X_C = x_C) P(X_B = x_B | X_C = x_C).
    \]
which expresses the matrix $   {\rm Flat}(A,B | C, x_C)(P)/  P(X_C = x_C) $
as a rank one matrix provided $P(X_C)$ is not zero.  So clearing denominators shows
that $   {\rm Flat}(A,B | C, x_C)(P)$ is  a rank one matrix (or the zero matrix if 
$P(X_C = x_C) = 0$).
\end{proof}

\begin{cor}\label{cor:rankflat}
    Let $P$ be a probability distribution that satisfies the conditional
    independence statement $X_A \ind X_B | X_C$.  Then
    the flattening matrix $   {\rm Flat}(A,B)(P) $
    has rank $\leq \#\calr_C$.
\end{cor}

\begin{proof}
    We have that
    \[
      {\rm Flat}(A,B)(P)  = \sum_{x_C \in \calr_C}    {\rm Flat}(A,B | C, x_C)(P) .
    \]
    Since each $   {\rm Flat}(A,B | C, x_C)(P) $ has rank $\leq 1$, and the
    rank is subadditive, we see that 
    \[ \rank {\rm Flat}(A,B)(P) \leq \#\calr_C .   \qedhere \]
\end{proof}

These statements hold in arbitrary DAG models, without reference to
the models arising from a network.  In the network models, we can use this
result to deduce rank conditions on flattening matrices when we 
cut edges in the network.   This follows from a simple observation.

\begin{thm}\label{thm:rankflat}
    Let $G$ be a DAG under the displayed tree model with
    an equivariant Markov model with $k$ state random variables.  
    Let $E$ be a set of
    edges of $G$ such that the removal of $E$ from $G$ disconnects 
    the graph into (at least) 2 subgraphs.  Let $A$ and $B$ be the two leaf sets of two
    disconnected components.  Then 
    \[
    \rank {\rm Flat}(A,B)(P) \leq k^{\#E}
    \]
    for any distribution $P$ in the model.
\end{thm}

\begin{proof}
    First, we perform a local modification to $G$.  For each edge $a \to b$
    in $E$, we replace it with a path $a \to c \to b$ (for some new independent
    vertex $c$ for each edge).  Let $G'$ be the resulting graph. 
    According to Proposition \ref{prop:degree2vertex},
    these local modifications of $G$ and $G'$ yield the same family of probability
    distributions if all the intermediate $c$ vertices are hidden.  Let $C$
    be the collection of all of those $c$ vertices that were added into $G'$,
    and $A$ and $B$ the two sets of leaves that are disconnected by removing the
    edge set $E$.  

    Since $A$ and $B$ are disconnected by removing the edges $E$ in $G$,
    this means that in $G'$, every chain from an $a \in A$ to a $b \in B$
    must pass through a vertex $c \in C$.  However, since each $c \in C$,
    only appears on a path $a' \rightarrow c \rightarrow b'$ in $G'$,
    this chain is blocked by $C$.  This implies that $X_A \ind X_B | X_C$
    in $G'$.  By Corollary \ref{cor:rankflat}, $\rank {\rm Flat}(A,B)(P) \leq k^{\#E}$.
    \end{proof}

\begin{ex}
Consider the $6$-sunlet network in Figure \ref{fig:6sunlet}.  
Removing the two edges $2 \rightarrow 1$ and $4 \rightarrow 5$
separates the leaves into two groups $A = \{\alpha, \zeta, \epsilon \}$
and $B = \{\beta, \gamma, \delta\}$.
So the flattening matrix ${\rm Flat}(A,B)(P)$ has
\[
\rank {\rm Flat}(A,B)(P)  \leq  k^2
\]
for a $k$ state phylogenetic model.  Note that ${\rm Flat}(A,B)(P) $ is a $k^3 \times k^3$
matrix in this case, so this rank condition is giving a nontrivial restriction on the matrix  ${\rm Flat}(A,B)(P) $.   
\end{ex}

As a complement to Theorem \ref{thm:rankflat}, results in the
literature on ranks of flattenings of tensors associated to phylogenetic
trees show in some situations that we can derive 
lower bounds on the rank of a flattening.
To explain this we need the notion of the parsimony score of a 
split $A|B$ which may not be displayed by a particular tree.  

\begin{defn}
    Let $T$ be a tree and let $A|B$ be a split of the leaves.  A labeling
    of the vertices of $T$ with $\{0,1\}$ is compatible with the split $A|B$ if 
    every $a \in A$ gets label $0$ and every $b \in B$ gets label $1$.
    The parsimony score of $A|B$ on $T$ is the smallest number of $0/1$
    edges in any compatible labeling of $T$.  Denote this by $\ell_T(A|B)$.
\end{defn}

Note that the split $A|B$ appears in the tree $T$ if and only if $\ell_T(A|B) = 1$. 

\begin{thm}\label{thm:ranklower}
    Let $G$ be a DAG evolving under the displayed tree model
    for random variables with $k$ states.  Let $A|B$
    be a partition of the leaves of $G$.  Suppose that one of the displayed trees $T$
    of $G$ has parsimony score $\ell_T(A|B)$.  Then for a generic probability distribution
    $P$ in the model
    \[
    \rank {\rm Flat}(A,B)(P)  \geq  \min(  k^{\#A}, k^{\#B}, k^{\ell_T(A|B)} ). 
    \]
\end{thm}

\begin{proof}
    The main feature we use is that the rank of a matrix is upper-semicontinuous, which implies that
    if there is a value of the parameters that achieves a particular rank, then generic
    points will also have rank at least that value.  The probability
    distributions for a fixed displayed tree $T$ always arise  in the closure
    of the displayed tree model (e.g. by setting all the reticulation parameters
    associated to the edges in that tree to $1$, and all other reticulation
    parameters to zero).  So it suffices to show that 
    the distribution for a tree with  the fixed parsimony score $\ell_T(A|B)$
    gives the appropriate rank.  However, this result for trees is show in \cite[Prop. 3.1]{Casanellas2011}
    and \cite[Thm. 8]{Snyman2023}.
\end{proof}

\begin{ex}
    Consider the $6$-sunlet network from the left of  Figure \ref{fig:6sunlet}
    and consider the split of the leaves $A|B = \{\alpha, \gamma, \epsilon\}|\{\beta, \delta, \zeta\} $.
    In both of the two displayed trees for this network, the parsimony
    score is  $\ell_T(A|B) = 3$ (which is straightforward to see since the labeling alternates as it goes around the sunlet).  So this shows that for
    a generic $P$ from  the network model
    \[
    \rank {\rm } {\rm Flat}(A,B)(P)  \geq  k^3
    \]
    by Theorem \ref{thm:ranklower}.  However, ${\rm Flat}(A,B)(P)$  is a $k^3 \times k^3$
    matrix, so the rank is generically equal to $k^3$.
\end{ex}

\section*{Acknowledgments}
This material is based upon work supported by the National Science Foundation under Grant No.~DMS-1929284 while the author was in residence at the Institute for Computational and Experimental Research in Mathematics in Providence, RI, during the semester program on
``Theory, Methods, and Applications of Quantitative Phylogenomics''.  The author thanks
Marta Casanellas, Aviva Englander, Jes\'us Fern\'andez-S\'anchez, Martin Frohn, Elizabeth Gross, Ben Hollering, Niels Holtgrefe, Mark Jones, and Leo van Iersel, as this project
arose as an offshoot to discussions around the related projects \cite{Casanellas2025, Englander2025}.

\bibliographystyle{amsplain}
\bibliography{dag.bib}

\end{document}